\documentclass[useAMS,usenatbib]{mn2e} 
\pdfoutput=1 
\usepackage{journals}
\usepackage[pdftex]{graphicx,color} 
\usepackage[latin1]{inputenc}
\usepackage{graphics} 
\usepackage{amsfonts} 
\usepackage{amsmath}
\usepackage{multicol} 
\usepackage{layout} 
\usepackage{amssymb}
\usepackage{color}
\title[Mass accretion history in cDE models] {Characterizing dark
  interactions with the halo mass accretion history and structural
  properties} \author[Giocoli et al. 2012] {\parbox{\textwidth}{Carlo
    Giocoli$^{1,2,3}$\thanks{E-mail:
      carlo.giocoli@unibo.it},      
    Federico Marulli$^{1,2,3}$, Marco Baldi$^{1}$, Lauro
    Moscardini$^{1,2,3}$, R. Benton Metcalf$^1$}
  \\ \\ 
  $^{1}$ Dipartimento di Fisica e
  Astronomia, Universit\`a di Bologna, viale Berti Pichat 6/2,
  40127, Bologna, Italy \\ 
  $^{2}$ INAF - Osservatorio Astronomico di
  Bologna, via Ranzani 1, 40127, Bologna, Italy \\ 
  $^{3}$ INFN - Sezione di Bologna, viale Berti Pichat 6/2, 
  40127, Bologna, Italy
}
\begin{document}
\date{}
\maketitle
\label{firstpage}
\pagerange{\pageref{firstpage}--\pageref{lastpage}} \pubyear{2012}
\begin{abstract}
  We study the  halo mass accretion history (MAH)  and its correlation
  with the internal structural properties in coupled dark energy (cDE)
  cosmologies. To accurately predict all the non-linear effects caused
  by dark  interactions, we use  the COupled Dark  Energy Cosmological
  Simulations ({\small CoDECS}).  We measure the halo concentration at
  $z=0$  and  the number  of  substructures  above a  mass  resolution
  threshold for  each halo.   Tracing the  halo merging  history trees
  back in time, following the mass of the main halo, we develope a MAH
  model  that  accurately  reproduces  the  halo  growth  in  term  of
  $M_{200}$    in   the    $\mathrm{\Lambda}$    cold   dark    matter
  ($\mathrm{\Lambda}$CDM  Universe;   we  then  compare  the   MAH  in
  different  cosmological  scenarios.   For  cDE models  with  a  weak
  constant  coupling,  our  MAH  model can  reproduce  the  simulation
  results,  within  $10\%$  of  accuracy, by  suitably  rescaling  the
  normalization  of  the  linear   matter  power  spectrum  at  $z=0$,
  $\sigma_{8}$.   However,  this is  not  the  case for  more  complex
  scenarios, like the  ``bouncing" cDE model, for  which the numerical
  analysis  shows a  rapid growth  of haloes  at high  redshifts, that
  cannot be reproduced by simply  rescaling the value of $\sigma_{8}$.
  Moreover,  at fixed  value  of $\sigma_{8}$,  $\mathrm{\Lambda}$cold
  dark  matter CDM  haloes  in these  cDE scenarios  tend  to be  more
  concentrated and have a larger  amount of substructures with respect
  to  $\mathrm{\Lambda}$CDM  predictions.    Finally,  we  present  an
  accurate model  that relates the  halo concentration to the  time at
  which it assembles  half or $4\%$ of its mass.   Combining this with
  our MAH model, we show  how halo concentrations change while varying
  only $\sigma_8$  in a $\mathrm{\Lambda}$CDM Universe,  at fixed halo
  mass.
\end{abstract}

\section{Introduction}
Understanding  the  formation  and  evolution  of  structures  in  the
Universe is  one of  the main goals  of present  cosmological studies.
Following the standard scenario, the formation of cosmic structures up
to protogalactic scale is due  to the gravitational instability of the
dark   matter   (DM)   \citep{frenk83,  davis85,   white88,   frenk90,
  springel05b,  springel08b}.  When  a density  fluctuation exceeds  a
certain value,  it collapses forming  a so-called DM halo.   The small
systems collapse  first in a  denser Universe and then  merge together
forming the larger ones, thereby giving rise to a hierarchical process
of structure formation.  In this scenario, galaxy clusters are located
at  the top  of the  merger history  pyramid and  today represent  the
largest virialized objects in the Universe.  The formation of luminous
objects happens  when baryons, feeling  the potential wells of  the DM
haloes,  fall inside  them  shocking, cooling  and eventually  forming
stars \citep{white78}.  New supplies of gas and galaxy mergers tend to
modify  the  dynamical and  morphological  properties  of the  forming
systems, and are closely linked to the mass accretion histories of the
haloes  they  inhabit.  A  detailed  understanding  of how  this  mass
accretion occurs  and how individual  halo properties depend  on their
merger histories  is of  fundamental importance for  predicting galaxy
properties within  the cold dark  matter (CDM) theory  and, similarly,
for using  observed galaxy  properties (as  e.g.  rotation  curves) to
test the paradigm.

Different definitions have been adopted in the literature to study the
halo growth  along the cosmic time  and its correlation with  the halo
clustering  on  large  scales  \citep{gao04,  gao07}.  From  different
analyses it  has emerged that while  the redshit, $z_f$, at  which the
main halo progenitor assembles a fraction, $f$, of its mass correlates
with its global structural properties (as e.g. its concentration, spin
parameter, subhalo  population, etc.),  the redshift, $z_c$,  at which
the main  halo progenitor  assembles a  constant central  mass, $M_c$,
mainly correlates with the typical formation  time of stars in a halo.
Considering  the  haloes  at   $z=0$  in  the  Millennium  Simulation,
\citet{li08}  have shown  that, while  $z_f$ decreases  with the  halo
mass,  $z_c$ grows  with it,  in agreement  with the  fact that  older
stellar populations tend to reside in more massive systems.

Many studies conducted on standard $\Lambda $CDM simulations have also
underlined  how structural  halo  properties,  like concentration  and
subhalo      population,     are      related      to     the      MAH
\citep{vandenbosch02,gao04,gao08}.   Less  massive  haloes,  typically
assembling a  given fraction of their  mass earlier, tend to  host few
substructures   than   the   more  massive   ones.    In   particular,
\citet{giocoli08b} found that  at a fixed halo  mass more concentrated
haloes  tend  to possess  few  substructures  because form  at  higher
redshift than the less concentrated  ones.  Extending these results to
the framework of the assembly bias we  would expect -- at a fixed halo
mass -- more concentrated, more relaxed, and less substructured haloes
to be on average more biased with respect to the DM.

Analytical models  of the DM  density distribution, based on  the halo
model \citep{seljak00, cooray02, giocoli10b}, require the knowledge of
the  halo mass  function, density  profile, concentration  and subhalo
population; as does the halo occupation distribution (HOD) approach to
describe the galaxy and the  quasar luminosity function and their bias
\citep{moster10, shen10, cacciato12}. At  the same time, these results
are useful to model the weak and strong lensing signals by large scale
structures  \citep{bartelmann01}  and  clusters  \citep{giocoli12}  in
standard and non-standard cosmologies.  Since many cosmological models
have  been proposed  as a  possible alternative  to the  standard {\em
  concordance}  $\Lambda$CDM scenario,  it is  natural to  investigate
whether   these  results   can  also   be  extended   to  non-standard
cosmologies.  In this paper, we will  focus on a range of DE scenarios
characterized by a non-vanishing coupling between the DE field and CDM
particles,  which go  under  the  name of  coupled  dark energy  (cDE)
\citep[][]{Wetterich_1995, Amendola_2000, Amendola_2004}.

This paper is divided in two  parts. In the first part, we investigate
the impact  of a number of  cDE models on the  formation and accretion
histories of  CDM haloes (\S \ref{secmodel}  -- \S \ref{secMAHmodel}).
To tackle this  point, we make use of two  theoretical techniques, one
analytic and one numeric. The former  one allows us to predict the MAH
through a  generalized version  of the \citet{press74}  formalism. The
numerical  approach aimed  at properly  describing all  the non-linear
effects at work. Here, we  consider the mock halo catalogues extracted
from the {\small CoDECS} simulations  and compare them to our analytic
MAH   predictions.    In   the   second  part   of   the   paper   (\S
\ref{secHaloProp}), we  exploit both the above  theoretical methods to
predict  the internal  structural properties  of CDM  haloes in  these
cosmological   scenarios.   In   particular,   we   derive  the   halo
concentration-mass relation, that is shown to provide a key observable
to discriminate between cDE cosmologies  and, in some cases, to remove
the degeneracy with the normalization of the power spectrum.

After a general  introduction to the cDE cosmologies  analysed in this
work  in   \S  \ref{secmodel},   we  describe  our   numerical  N-body
simulations in  \S \ref{secnumsim}.  Details on the  method we  use to
climb   the   halo   merging   history   trees   are   given   in   \S
\ref{secMergerTree},   while   our   theoretical   predictions,   both
analytical  and  numerical,  on  the  halo  MAH  are  provided  in  \S
\ref{secMAHmodel}.  In  \S \ref{secHaloProp}, we investigate  the halo
internal structural  properties.  Finally, in  \S \ref{secConclusions}
we draw our conclusions.

\section{The cDE models}
\label{secmodel}
cDE  models  have been  proposed  as  a  possible alternative  to  the
cosmological constant $\Lambda$ and to standard uncoupled Quintessence
models \citep[][]{Wetterich_1988,  Ratra_Peebles_1988} to  explain the
observed      accelerated      expansion     of      the      Universe
\citep[][]{Riess_etal_1998,  Perlmutter_etal_1999, Schmidt_etal_1998},
as they alleviate  some of the fine-tuning  problems that characterize
the  latter scenarios.   In order  to be  viable, cDE  models need  to
assume a negligible interaction of  the DE field to baryonic particles
\cite[][]{Damour_Gibbons_Gundlach_1990},    since    the    long-range
fifth-force  that   arises  between   coupled  particle  pairs   as  a
consequence  of the  interaction  with the  DE  field would  otherwise
violate solar-system constraints  on scalar-tensor theories \citep[see
  e.g.][]{Bertotti_Iess_Tortora_2003,  Will_2005}.    Consequently,  a
large  number of  cDE  models characterized  by  a direct  interaction
between DE  and CDM \citep[see  e.g.][]{Wetterich_1995, Amendola_2000,
  Amendola_2004,  Farrar2004,   CalderaCabral_2009,  Koyama_etal_2009,
  Baldi_2011a} or massive neutrinos \citep[see e.g.][]{Wetterich_2007,
  Amendola_Baldi_Wetterich_2008} have  been proposed in  recent years.
The  basic properties  of cDE  models  and their  impact on  structure
formation have  been thoroughly  discussed in  the literature.   For a
self-consistent introduction  on cDE scenarios we  suggest for example
the   recent   reviews   \citet[][-   Section   1.4.3]{Tsujikawa_2010,
  DeFelice_Tsujikawa_2010, Euclid_TWG}.   For the aims of  the present
work,  it   is  sufficient  to   summarize  the  main   features  that
characterize cDE models  in general, and the  specific realizations of
the cDE scenario that will be investigated here.

In general, cDE models are characterized by two free functions that
fully determine the background evolution of the Universe and the
linear and non-linear growth of density perturbations.  These are the
self-interaction potential $V(\phi )$ and the coupling function $\beta
(\phi )$, where $\phi $ is a classical scalar field playing the role
of the cosmic DE.  The coupling function $\beta (\phi )$ determines
the strength of the interaction between DE and CDM (in the case
investigated here) and directly affects the evolution of density
perturbations through: $i)$ a long-range attractive fifth-force of
order $\beta ^{2}$ times the gravitational strength, and $ii)$ a
velocity-dependent acceleration on coupled CDM particles proportional
to $\dot{\phi }\beta $ \citep[see e.g.][]{Baldi_2011b}. Differently
from the fifth-force term, which is always attractive regardless of
the sign of the coupling function and of the dynamical evolution of
the DE scalar field, the friction term can take both positive and
negative signs depending on the relative signs of the scalar field
velocity $\dot{\phi }$ and the coupling function $\beta (\phi )$.
Such feature can have very significant consequences on the evolution
of structure formation in case the friction term changes sign during
the cosmic evolution \citep[][]{Baldi_2011c,Tarrant_etal_2012} as for
the case of the ``bouncing" cDE model \citep[][]{Baldi_2011c}
investigated in this work.
\begin{table*}
\caption{The cDE models of the {\small CoDECS} suite considered in the
  present  work. All  models  have the  same  normalization of  scalar
  perturbations at  $z_{\rm CMB}\approx  1100$ leading to  a different
  value of $\sigma  _{8}$. Besides $\sigma _{8}$ and the  value of the
  DE equation of state $w_{\phi }$ all the models share the same WMAP7
  \citep[][]{wmap7} cosmological parameters at the present time.}
\label{tab:models}
\begin{tabular}{llcccccccc}
\hline
\hline
Model & Potential  &  
$\alpha $ &
$\beta _{0}$ &
$\beta _{1}$ &
\begin{minipage}{45pt}
Scalar field \\ normalization
\end{minipage} &
\begin{minipage}{45pt}
Potential \\ normalization
\end{minipage} &
$w_{\phi }(z=0)$ &
${\cal A}_{s}(z_{\rm CMB})$ &
$\sigma _{8}(z=0)$\\
\\
\hline
$\Lambda $CDM & $V(\phi ) = A$ & -- & -- & -- & -- & $A = 0.0219$ & $-1.0$ & $2.42 \times 10^{-9}$ & $0.809$ \\
EXP003 & $V(\phi ) = Ae^{-\alpha \phi }$  & 0.08 & 0.15 & 0 & $\phi (z=0) = 0$ & $A=0.0218$ & $-0.992$ & $2.42 \times 10^{-9}$ & $0.967$\\
EXP008e3 & $V(\phi ) = Ae^{-\alpha \phi }$  & 0.08 & 0.4 & 3 & $\phi (z=0) = 0$ & $A=0.0217$ & $-0.982$ & $2.42 \times 10^{-9}$ & $0.895$ \\
SUGRA003 & $V(\phi ) = A\phi ^{-\alpha }e^{\phi ^{2}/2}$  & 2.15 & -0.15 & 0 & $\phi (z\rightarrow \infty ) = \sqrt{\alpha }$ & $A=0.0202$ & $-0.901$ & $2.42 \times 10^{-9}$ & $0.806$ \\
\hline
\hline
\end{tabular}
\end{table*}

\section{Numerical Simulations}
\label{secnumsim}
We make use  of the public halo and subhalo  catalogues of the {\small
  CoDECS}\footnote{www.marcobaldi.it/CoDECS}               simulations
\citep[][]{CoDECS}  --  the  largest   suite  of  cosmological  N-body
simulations of cDE models to date -- to follow the accretion histories
of CDM haloes by building the  full merger trees of all the structures
identified at $z=0$  up to $z=60$.  In our analysis,  we will consider
the  different combinations  of the  potential and  coupling functions
that are available within the  {\small CoDECS} suite of cDE scenarios,
defined by the following general expressions:
\begin{itemize}
\item {\em Exponential potential} \citep[][]{Lucchin_Matarrese_1984,Wetterich_1988}:
\begin{equation}
\label{exponential}
V(\phi ) = Ae^{-\alpha \phi } \,;
\end{equation}
\item {\em SUGRA potential} \citep[][]{Brax_Martin_1999}:
\begin{equation}
\label{sugra}
V(\phi ) = A\phi ^{-\alpha }e^{\phi ^{2}/2}\,;
\end{equation}
\item {\em Constant coupling} \citep[][]{Wetterich_1995,Amendola_2000}:
\begin{equation}
\label{constant_c}
\beta (\phi ) = \beta _{0} = {\rm const.} \,;
\end{equation}
\item {\em Time-dependent coupling} \citep[][]{Amendola_2004,Baldi_2011a}:
\begin{equation}
\label{variable_c}
\beta (\phi ) = \beta _{0}e^{\beta _{1}\phi }\,.
\end{equation}
\end{itemize}
In  particular,  while  standard   cDE  models  are  characterized  by
``runaway"  potentials  (like  e.g.    the  exponential  potential  of
equation~\ref{exponential})  and a  constant  coupling, time-dependent  cDE
models feature  the same type  of potentials with a  coupling function
that evolves with  the scalar field itself.   Finally, the combination
of   a   ``confining"   potential   like  the   SUGRA   potential   of
equation.~\ref{sugra} with a  negative constant coupling characterizes the
so-called ``bouncing" cDE scenario \citep[][]{Baldi_2011c}.  A summary
of the cDE models that are  investigated in the present work, with the
corresponding parameters, is given in Table~\ref{tab:models}.
For such scenarios, we will make use of the publicly-available data of
the {\small L-CoDECS} series to derive halo accretion histories in the
different cosmologies.

The  {\small  L-CoDECS}  runs  are  collisionless  N-body  simulations
carried  out on  a periodic  cosmological  box of  $1$ Gpc$/h$  aside,
filled with  $2\times 1024^{3}$ equally  sampling the coupled  CDM and
the uncoupled  baryon fluids.  The  baryons are treated as  a separate
family of  collisionless particles,  and no hydrodynamic  treatment is
included in the simulations. The mass  resolution at $z=0$ is $m_{c} =
5.84\times  10^{10}$ M$_{\odot}/h$  and $m_{b}  = 1.17\times  10^{10}$
M$_{\odot}/h$ for  CDM and  baryon particles, respectively,  while the
gravitational  softening is  $\epsilon  _{g} =  20$  kpc$/h$. All  the
models virtually share the same  initial conditions at the redshift of
last  scattering   $z_{\rm  ls}\approx  1100$  and   are  consequently
characterized  by a  different value  of $\sigma  _{8}$, due  to their
different      growth      histories       (as      summarized      in
Table~\ref{tab:models}). Except  for the  different values  of $\sigma
_{8}$ and  of the DE equation  of state parameter $w_{\phi}$,  all the
models  share the  same cosmological  parameters at  $z=0$, consistent
with    the    WMAP7    results    (\citet[][]{wmap7},    see    Table
\ref{tab:parameters}).   This   provides  a  self-consistent   set  of
cosmologies that can  be directly compared with each  other and tested
with present and  future observations.  The viability  of these models
in terms  of CMB observables has  yet to be properly  investigated, in
particular  for  what concerns  the  impact  of variable-coupling  and
``bouncing" cDE models  on the large-scale power  of CMB anisotropies.
Although such  analysis might possibly  lead to tighter bounds  on the
coupling and on  the potential functions than the ones  allowed in the
present  work,   here  we  are  mainly   interested  in  qualitatively
understanding the impact of cDE  scenarios on the formation history of
CDM haloes  at late  times, and we  deliberately choose  quite extreme
values of the models parameters in order to maximize the effects under
investigation.

The {\small  CoDECS} simulations  have already  been used  for several
investigations.  In particular, \citet{Lee_Baldi_2011} exploited these
data  to  study  the  pairwise infall  velocity  of  colliding  galaxy
clusters  in  cDE  models,  demonstrating  that  DE  interactions  can
significantly  enhance the  probability  of high-velocity  collisions.
\citet{Beynon_etal_2012}  provided  forecasts  for  the  weak  lensing
constraining  power  of  future galaxy  surveys,  while  \citet{cui12}
exploited the  same data  to test  the universality  of the  halo mass
function. Finally,  the {\small CoDECS}  data have been used  to study
how   DE   interactions  modify   the   halo   clustering,  bias   and
redshift-space distortions  \citep{Marulli_Baldi_Moscardini_2012}, and
how   they    can   shift    the   baryonic    acoustic   oscillations
\citep{vera2012}.

\begin{table}
\begin{center}
\caption{The set of cosmological parameters assumed for all the models
  included in  the {\small  CoDECS} Project,  consistent with  the $7$
  year  results  of   the  WMAP  collaboration  for   CMB  data  alone
  \citep[][]{wmap7}.}
\label{tab:parameters}
\begin{tabular}{cc}
\hline
Parameter & Value\\
\hline
$H_{0}$ & 70.3 km s$^{-1}$ Mpc$^{-1}$\\
$\Omega _{\rm CDM} $ & 0.226 \\
$\Omega _{\rm DE} $ & 0.729 \\
$\sigma_{8}$ & 0.809\\
$ \Omega _{b} $ & 0.0451 \\
$n_{s}$ & 0.966\\
\hline
\end{tabular}
\end{center}
\end{table}

\section{The Halo Merger History Tree}
\label{secMergerTree}
For each simulation  snapshot, we identify haloes on the  fly by means
of a  Fried-of-Friend (FoF)  algorithm with linking  parameter $b=0.2$
times the  mean interparticle  separation.  Within  each FoF  group we
also   identify   gravitationally   bound  substructures   using   the
\textsc{subfind}   algorithm  \citep{springel01b}.    \textsc{subfind}
searches for  overdense regions within a  FoF group using a  local SPH
density  estimate,  identifying  substructure  candidates  as  regions
bounded by  an isodensity surface that  crosses a saddle point  of the
density  field,  and testing  that  these  possible substructures  are
physically bound with an iterative  unbinding procedure.  For both FoF
and \textsc{subfind} catalogues we select  and store systems with more
than 20  particles, and define  their centres  as the position  of the
particle with  the minimum  gravitational potential.   It is  worth to
notice that while  the subhaloes have a well defined  mass that is the
sum of  the mass of  all particles  belonging to them,  different mass
definitions  are associated  with  the FoF  groups \citep{lukic09}.   We
define as $M_{FoF}$  the sum of the masses of  all particles belonging
to the FoF group, $M_{200}$ the mass around the FoF centre enclosing a
density that is $200$ times the critical one, and $M_{vir}$ as the one
enclosing the  virial overdensity $\Delta_{vir}$, as  predicted by the
spherical collapse model. We notice that in the $\Lambda $CDM Universe
at  $z=0$   generally  it   is  true   that  $M_{200}<M_{vir}<M_{FoF}$
\citep{eke96,bryan98}.

It  is interesting  to  notice  that the  definition  of $M_{vir}$  is
cosmology  dependent,  since  it  is  related  to  the  $\Delta_{vir}$
definition derived from the  spherical collapse model.  Generally this
quantity  is not  directly obtainable  for any  arbitrary DM  and dark
energy model.  For  example, \citet{pace10} have shown  how to compute
$\Delta_{vir}$  from  the  non-linear differential  equation  for  the
evolution  of  the  density  contrast   for  dynamical  and  early  DE
cosmologies.  For $\mathrm{\Lambda}$CDM  models many fitting functions
are available that depend on the  redshift evolution of the DM content
in the  Universe \citep{navarro97,bryan98},  while no  simple formulae
exist $\Delta_{vir}$ within the cDE  models investigated in this work.
Since  $M_{200}$ is  cosmology independent,  we will  adopt this  mass
definition in what follows.

For each  subhalo, starting  from redshift $z=0$,  we follow  back its
merger  history tree  by  requiring  to have  at  least one  descended
subhalo at  the previous snapshot \citep{boylan-kolchin09}.   In order
to trace  back in time the  growth in terms of  $M_{200}$, we consider
only  \emph{the  dominant subhalo}  of  each  FoF  group to  which  is
associated the  value of $M_{200}$ of  the group.  In this  process we
link  together  among  the  different simulation  snapshots  the  main
progenitor,  defined as  \emph{the dominant  subhalo} that  donate the
largest  number of  particles  between two  consecutive snapshots  and
\emph{the satellites}  that represent  the progenitor  haloes accreted
on to the main progenitor branch of the tree \citep{tormen04}.

\section{The Mass Accretion History}
\label{secMAHmodel}

Before  introducing the  results  of  the other  cDE  models, in  this
section, we review the model of the mass accretion history proposed by
\citet{giocoli12b} for a $\mathrm{\Lambda}$CDM  model and modify it to
be consistent  with the $M_{200}$  halo mass definition.   This simple
model  allows   us  to  derive  the   generalized  formation  redshift
distribution,  defined  as  the  redshift   at  which  the  main  halo
progenitor assembles a fraction $f$ of its $M_{200}$ mass at $z=0$, or
more  in general  at any  considered redshift  $z_0$. Then,  from such
distribution we  can compute the  halo mass accretion  history.  Since
haloes form  as a  consequence of gravitational  instability processes
that start within the initial DM  density field and then grow together
via  merging events,  a few  important consequences  are: (i)  the MAH
depends on  the initial  density fluctuation  field --  and so  on the
initial  matter power  spectrum;  (ii) it  also  depends on  cosmology
through the  background expansion  of the Universe,  (iii) at  a given
redshift more  massive haloes  grow faster  -- since  they sit  on the
highest peaks of the density fluctuation  field; (iv) at a given mass,
the  higher the  redshift  is the  higher the  halo  mass growth  rate
is. These  points will become more  clear along the discussion  of the
results presented in the following sections.

\subsection{Mass Accretion History in a $\mathrm{\Lambda}$CDM Universe: 
revisiting    the    \citet{giocoli12b}     model    for    $M_{200}$}
\citet{giocoli12b} developed  a simple and accurate  model to describe
the growth of CDM haloes along  cosmic time. Their study was conducted
analysing  the  merger  trees  extracted from  a  cosmological  N-body
simulation of  a $\mathrm{\Lambda CDM}$ Universe,  the GIF2 simulation
\citep{gao04,giocoli08b,giocoli10}.

\begin{figure}
\includegraphics[width=\hsize]{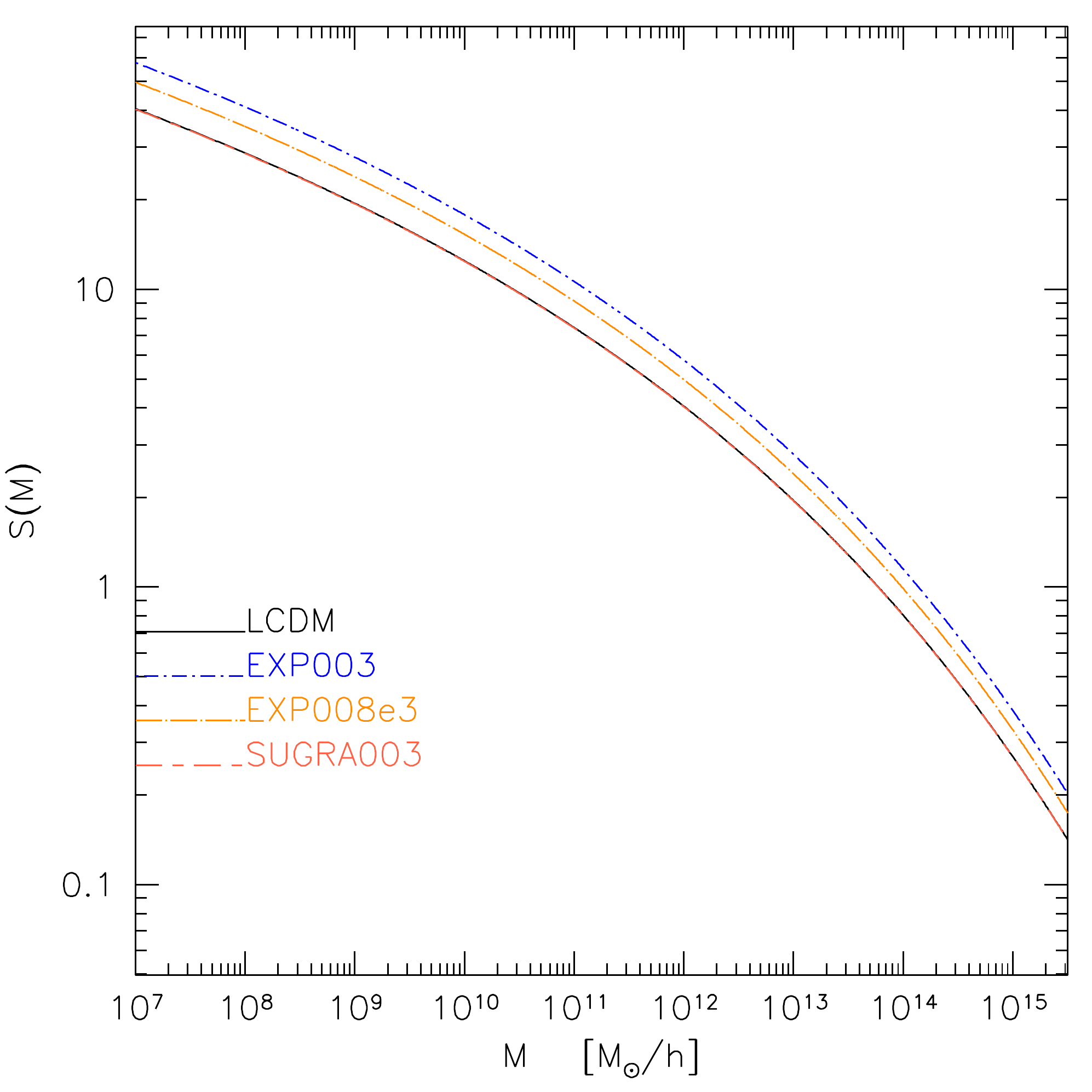}
\caption{The mass variance  $S$ (see equation~\ref{eqmassvariance}) as
  a function of  the halo mass. The different curves  show the results
  for the different  linear power spectra of the  models considered in
  this work.  Their  shapes are independent of  the cosmological model
  considered while  the normalizations change depending  on the linear
  power   spectrum  normalization   and   so   on  the   corresponding
  $\sigma_8$.\label{figSM}}
\end{figure}

Before describing  the model,  let us  first introduce  some universal
quantities that will  be used throughout this work.   We define $S(M)$
as  the variance  of linear  fluctuation  field when  smoothed with  a
top-hat filter on a scale $R=(3 M / 4 \pi \bar{\rho})^{1/3}$:
\begin{equation}
\label{eqmassvariance}
S(M) = \dfrac{1}{2 \pi^2} \int_0^{\infty} W^{2}(kR) P_{lin}(k) k^2 \mathrm{d} k\,,
\end{equation}
 where  $\bar{\rho}$  is  the  comoving  density  of  the  background,
 $P_{lin}(k)$ the linear matter power spectrum and $W(kR)$ the top-hat
 window  function.   We also  define  $\delta_{c}(z)$  as the  initial
 overdensity required for spherical collapse at redshift $z$:
\begin{equation}
\delta_{c}(z) = \dfrac{\delta_{c,lin}}{D_{+}(z)}\,,
\end{equation}
where $\delta_{c,lin}(z)$  is the  linear overdensity at  redshift $z$
and $D_{+}(z)$  the growth factor  normalized to unity at  the present
time.  For a CDM Universe,  \citet{nakamura97} have presented a useful
fitting  function for  the linear  overdensity  as a  function of  the
matter density parameter, $\Omega_m$, that can be read as:
\begin{equation}
\delta_{c,lin}(z) = 1.686 [ 1 + 0.123 \log (\Omega_m(z)] \,
\end{equation}
and that will be used throughout this paper.  In Fig.~\ref{figSM} we
show the mass variance for  the four cosmological models considered in
this  work. We  notice that  all of  them have  the same  shape but  a
different  normalization.  This  is  due to  the  different values  of
$\sigma_8$ defining the  variance of linear density  fluctuations on a
scale of $8$ Mpc/$h$:
\begin{equation}
\sigma_8^2(M) = \dfrac{1}{2 \pi^2} \int_0^{\infty} W^{2}(8k) P_{lin}(k) k^2 \mathrm{d} k\,;
\end{equation}
the  higher is  $\sigma_8$ the  higher the  normalization of  the mass
variance and viceversa.

Following  the  formalism  by   \citet{lacey93},  we  can  define  the
formation redshift $z_{f}$  of a halo of mass $M$  (at redshift $z_0$)
as the  redshift at which the  main halo progenitor assembles  for the
first time half of its  mass.  Their proposed redshift distribution at
half-mass, in terms of universal variables, can be read as:
\begin{equation}
p (w_f) = 2 w_f\, \mathrm{erfc} \left( \dfrac{w_f}{\sqrt{2}} \right),
\label{eqlacey}
\end{equation}
where $w_f  = (\delta_c(z_f)-\delta_c(z_0))  / \sqrt{S(M/2)  - S(M)}$.
It is  interesting to notice  that written  in this way  the formation
redshift distribution is independent both of the halo mass $M$ and the
redshift $z_0$. In a more  general way \citet{nusser99} have developed
a formalism to describe the redshift at which the main halo progenitor
assembles a fraction $1/2 \leq f<1$ of its mass, that is: \small
\begin{equation}
\
 p(w_f) = 2 w_f \, \left(\dfrac{1}{f} -1 \right) \mathrm{erfc}\left(
 \dfrac{w_f}{\sqrt{2}}\right) + \left( 2 - \frac{1}{f} \right)
 \sqrt{\dfrac{2}{\pi}} \mathrm{exp}\left( - \dfrac{w_f^2}{2}
 \right)\,,
\label{eqnusser}
\end{equation}
\normalsize  which  for  $f=1/2$   gives  back  the  above  expression
~(\ref{eqlacey}).    However,  comparing   these  distributions   with
measurements  performed on  numerical simulations,  \citet{giocoli07a}
have  shown that  they tend  to underestimate  the formation  redshift
distribution.  To reconcile  theory and  simulations they  proposed to
modify  $w_f  \rightarrow \sqrt{0.707}  w_f$  in  the context  of  the
ellipsoidal collapse model \citep{sheth99b,sheth01b,sheth02,despali12}.
\citet{giocoli12b}  have  shown  that  in  numerical  simulations  the
modified equation~(\ref{eqnusser}) works quite well for smaller values
of $f$.  They  also provided a new function to  describe the formation
redshift distribution for any fraction $0<f<1$:
\begin{equation}
p(w_f) = \dfrac{\alpha_f w \mathrm{e}^{w^2/2}}{\left[ \mathrm{e}^{w^2/2} + \alpha_f - 1\right]^2}\,,
\label{diffeqmodel1}
\end{equation}
where $\alpha _{f}$ represents a free parameter. By integrating
equation~\ref{diffeqmodel1} one gets the following cumulative generalized
formation redshift distribution:
\begin{equation}
 \label{eqmodel1}
 P(>w_f) = \frac{\alpha_f}{\mathrm{e}^{w_f^2/2}+\alpha_f-1}\,.
\end{equation}
Since the previous equation can be inverted, it is possible to
analytically compute the median value $w_f=\tilde{w}_{f}$ given by
definition when $P(>\tilde{w}_{f})=1/2$; this gives the median
redshift $z_f = \delta_c^{-1}(z_f)$ at which the main halo progenitor
assembles a fraction $f$ of its mass $M$:
\begin{equation}
z_f: \;\;\; \delta_c(z_f) = \delta_c(z_0) + \tilde{w}_{f}\, \sqrt{S(fM)-S(M)}\,,
\label{eqmahmodz}
\end{equation}
where
\begin{equation}
\tilde{w}_{f} = \sqrt{2\, \ln \left(\alpha_f + 1 \right)}\,.
 \label{eqmahmod1}
\end{equation} 

\begin{figure*}
\begin{center}
\includegraphics[width=7cm]{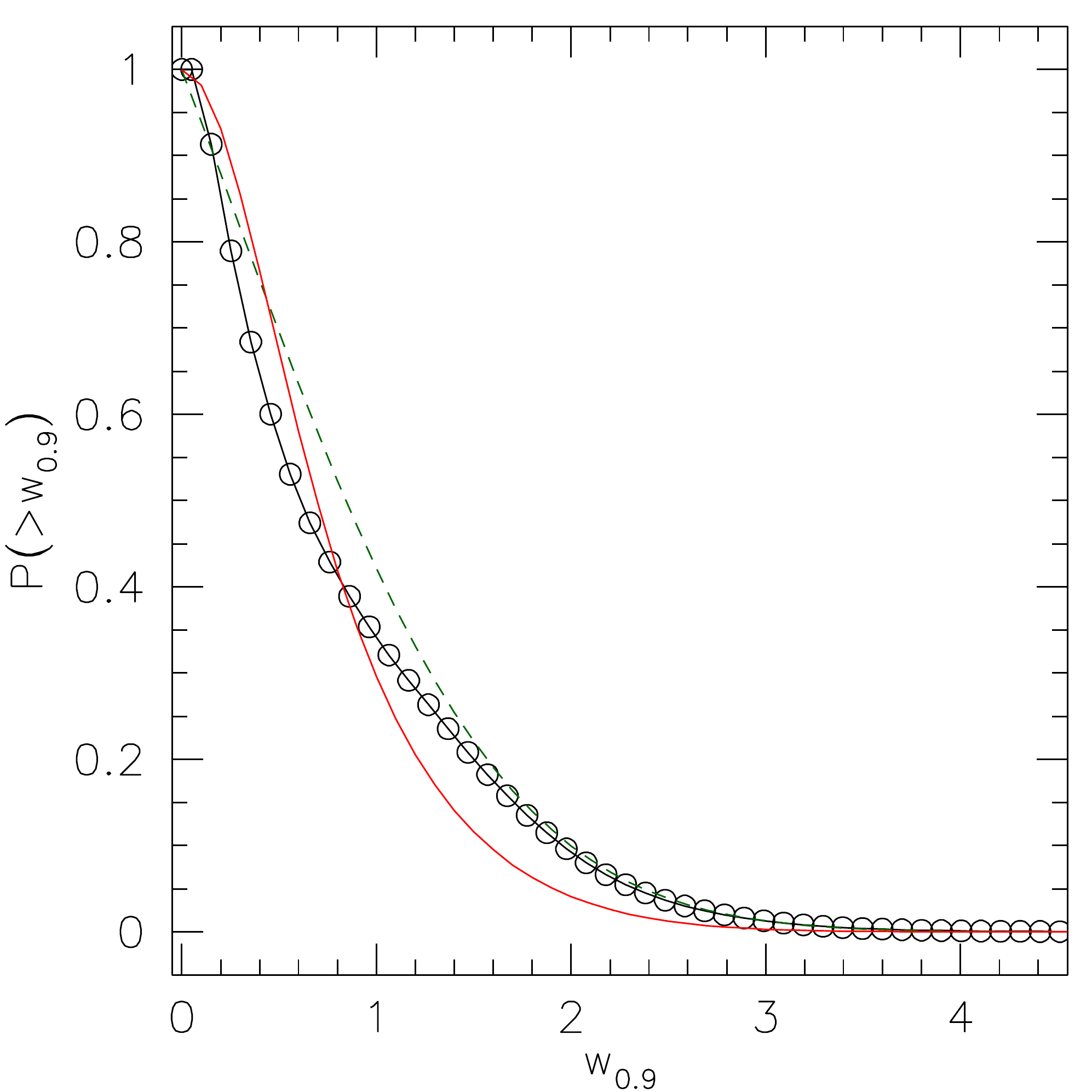}
\includegraphics[width=7cm]{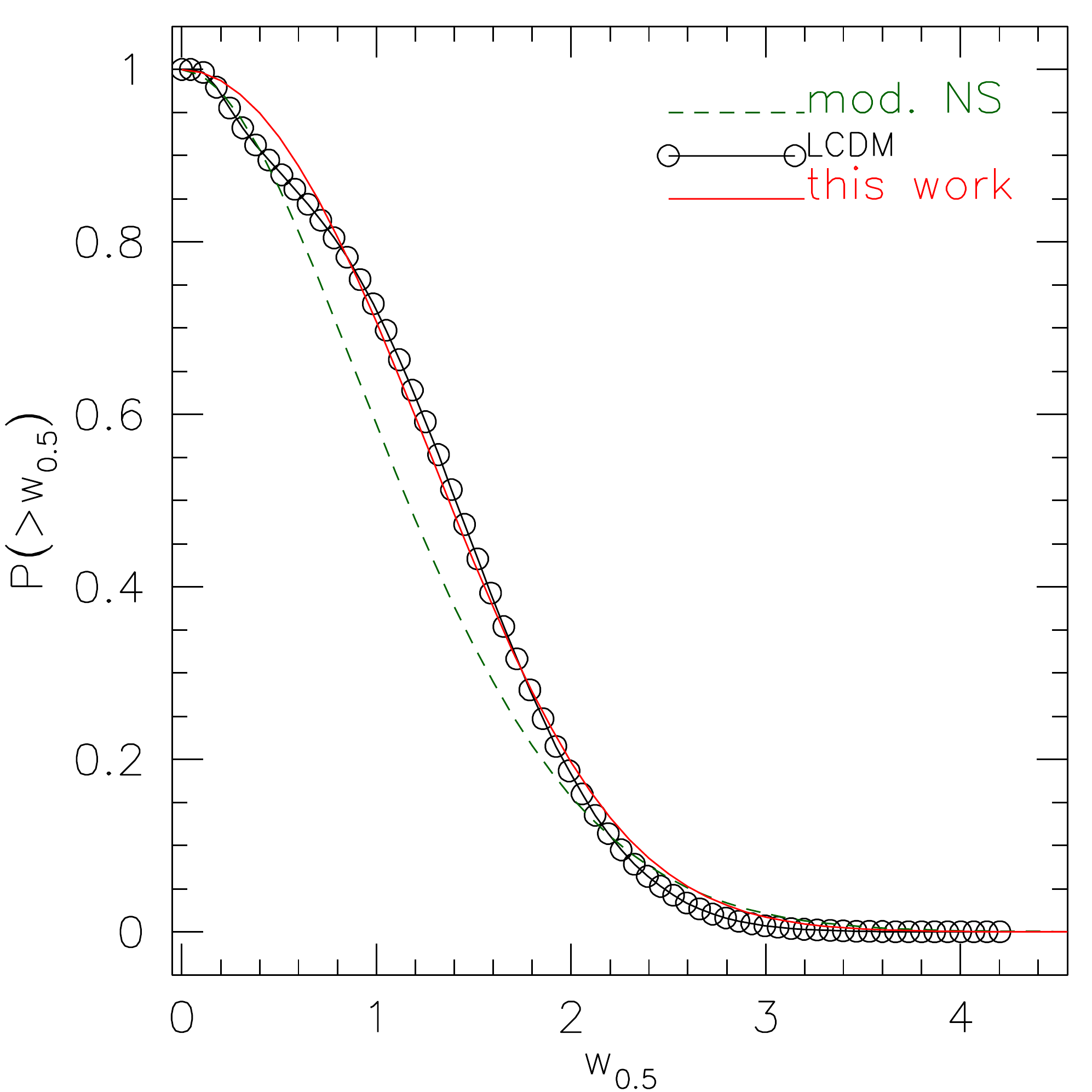}
\includegraphics[width=7cm]{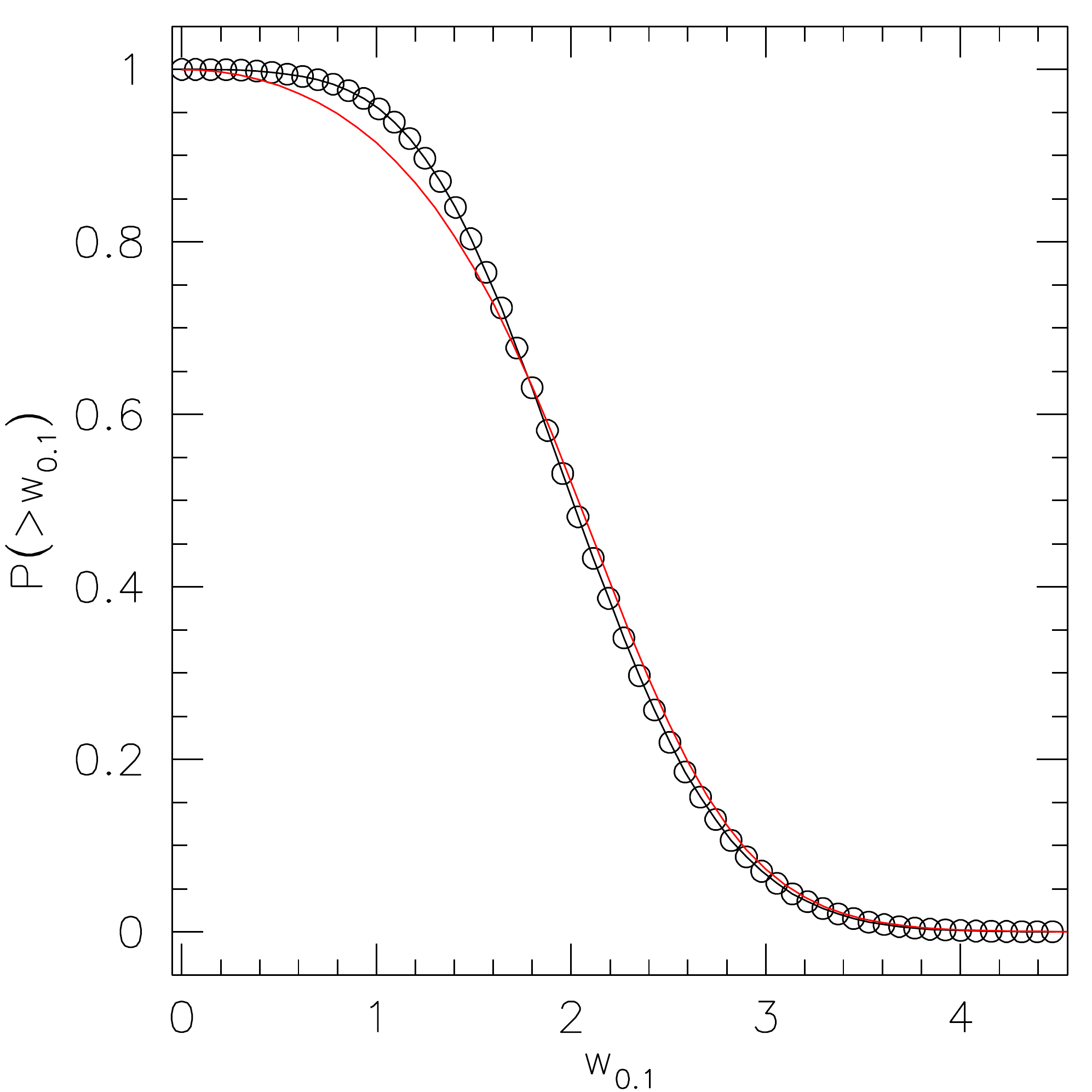}
\includegraphics[width=7cm]{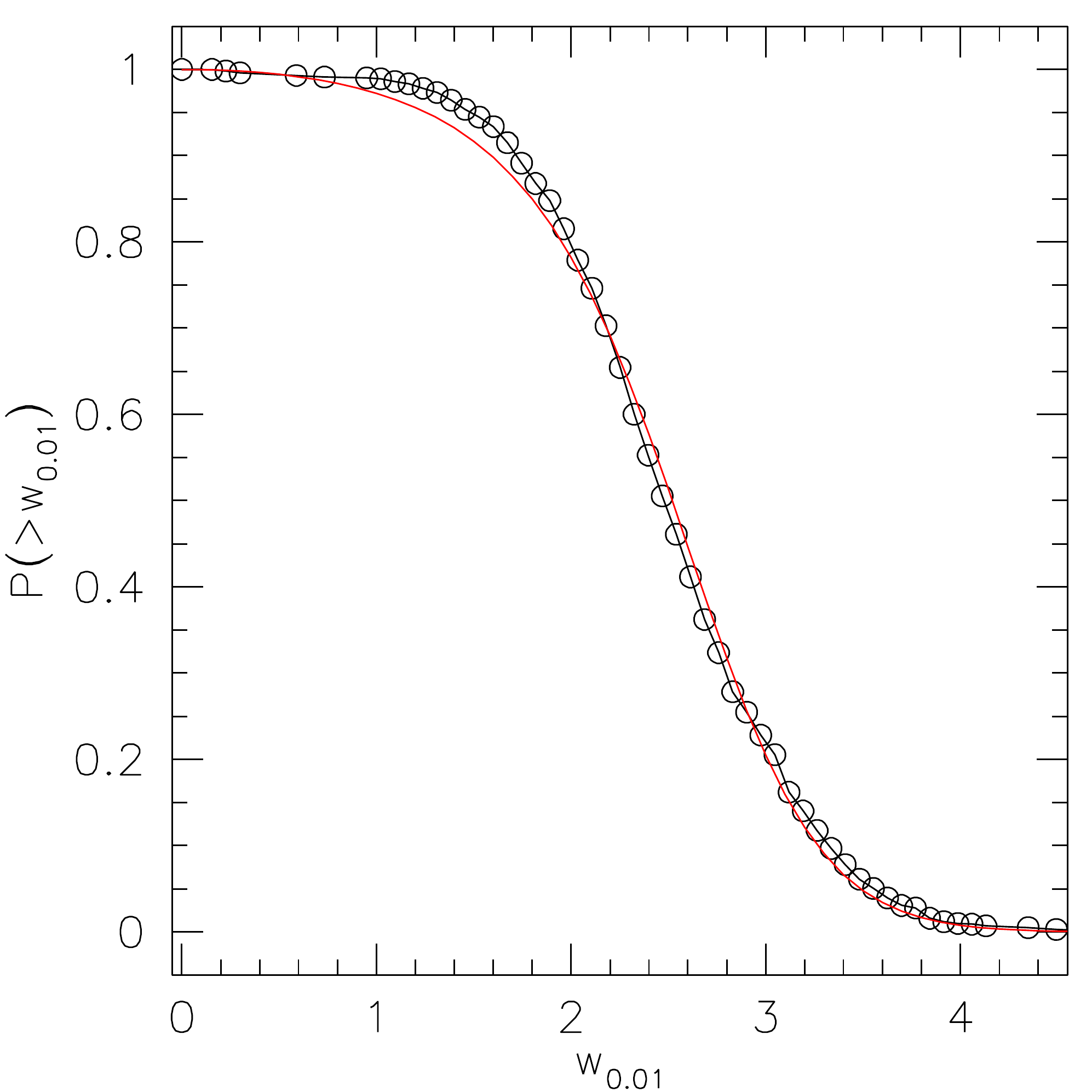}
\caption{Cumulative generalized formation  distribution, as a function
  of $w_f$, when  the main halo progenitor  assembles $90\%$ (top-left
  panel ),  $50\%$ (top-right  panel), $10\%$ (bottom-left  panel) and
  $1\%$  (bottom-right  panel)  of  its  $M_{200}$  mass  at  redshift
  zero. The  data points  show the  measurements for  the $\Lambda$CDM
  simulation considering all haloes with at least $40$ particles.  The
  dashed   curve  on   the  top   panels  refers   to  the   model  by
  \citet{nusser99} -- valid  only for $1/2<f \leq 1$,  while the solid
  curve  represents  equation~(\ref{eqmodel1}) with  the  best-fitting
  parameter $\alpha_f$.  \label{figcumPw} We  recall that  we consider
  only haloes  that along their  growth never exceed more  than $10\%$
  their mass at the present time.}
\end{center}
\end{figure*}

In Fig.~\ref{figcumPw} we  show the cumulative generalized formation
redshift distribution in terms of the rescaled variable $w_f$ when the
main halo  progenitors assemble  $90\%$, $50\%$,  $10\%$ and  $1\%$ of
their  $M_{200}$  mass  at  $z=0$.   The  data  points  represent  the
measurements in  the $\mathrm{\Lambda}$CDM simulation  considering all
haloes at  $z_0=0$ with  mass larger than  $M_{200}> 5  \times 10^{12}
M_{\odot}/h$ (which means that systems are resolved with at least $40$
particles) and  that never  exceed along  their growth  histories more
than  $10\%$ of  their  mass at  the present  time.   This ensures  to
consider  only  haloes that  grow  mainly  hierarchically and  do  not
fragment as a consequence of  violent merging events.  With respect to
\citet{giocoli12b},  in this  case haloes  are followed  back in  time
considering  $M_{200}$ as  mass  definition instead  of $M_{vir}$  The
solid curves  show the  best-fitting model of  equation~(\ref{eqmodel1}), while
the  dashed  curves  represent equation~(\ref{eqnusser})  modified  as
suggested by \citet{giocoli07a}.

From  the  figure we  notice  that  -- since  at  $z=0$  the value  of
$M_{200}$ is  smaller than that  of $M_{vir}$ -- following  the merger
tree back in  time in terms of  $M_{200}$ results in a  value of $z_f$
typically  higher than  that obtained  by following  back the  tree in
terms  of $M_{vir}$  \citep[see][or the  Appendix \ref{app1}  where we
  show  the   same  cumulative  distribution  of   formation  redshift
  following  the  haloes  in  term   of  their  virial  mass  for  the
  $\mathrm{\Lambda}$CDM run]{giocoli12b}.

\begin{figure}
\includegraphics[width=\hsize]{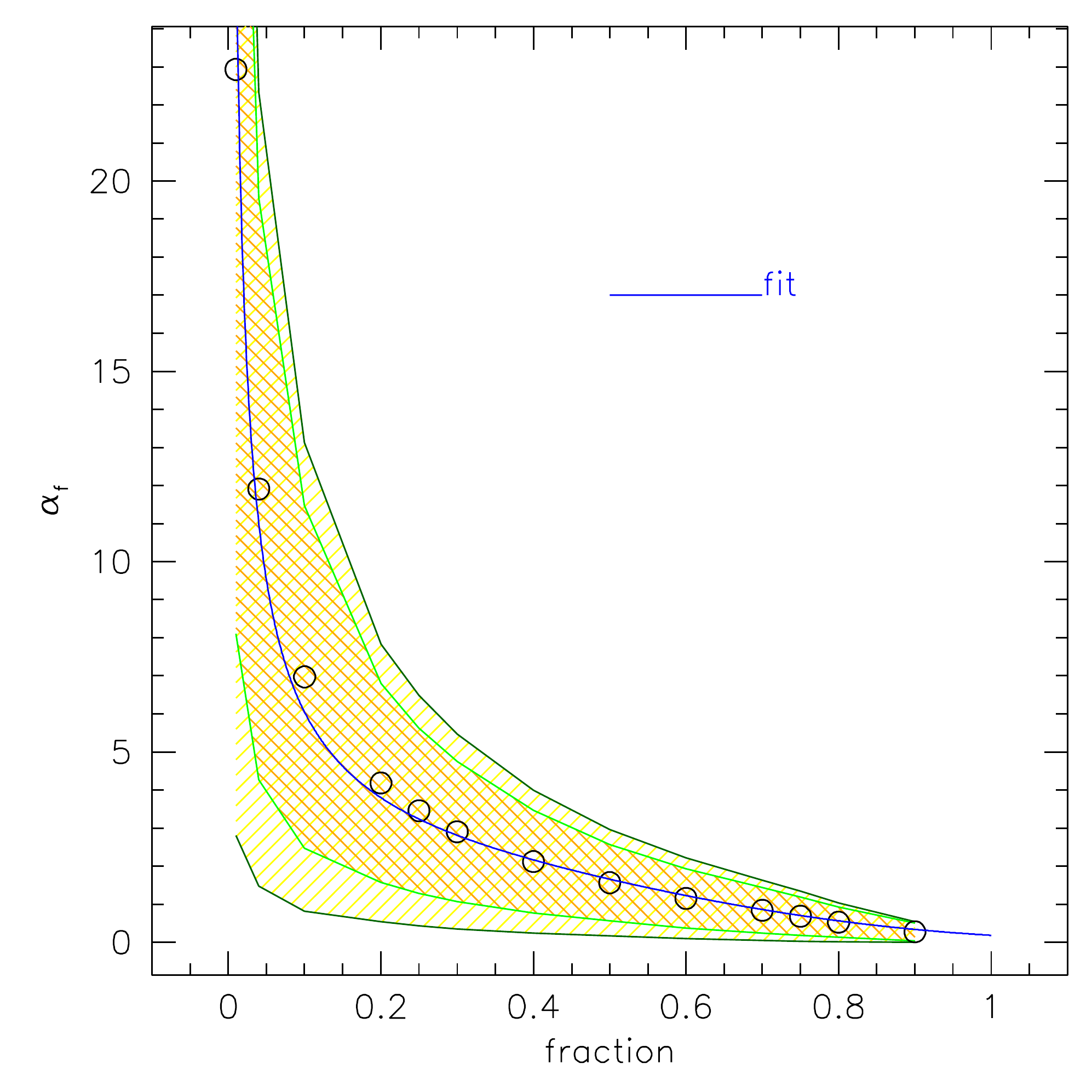}
\caption{Dependence   of    $\alpha_f$,   the   free    parameter   in
  equation~(\ref{eqmodel1}),    on   the    mass    fraction   $f    =
  M_{200}(z_f)/M_{200}(0)$  assumed  at  the  formation.   The  shaded
  regions represent  the 1$\sigma$  and 2$\sigma$ contours.   The blue
  curve represents the best-fitting relation. \label{figbestfit}}
\end{figure}

In  Fig.~\ref{figbestfit} we  show  the best-fitting  $\alpha_f$ as  a
function  of  $f$.   The  open  circles  show  the  best  fit  to  the
$\mathrm{\Lambda}$CDM simulation  measurement for different  values of
the assembled mass  fraction, while the solid line refers  to the best
fit to the data given by:
\begin{equation}
 \alpha_f = \dfrac{1.365}{f^{0.65}}\, \mathrm{e}^{-2 f^3}\,.
\end{equation}
Finally,  the shaded  regions  represent the  1$\sigma$ and  2$\sigma$
contours of $\Delta  \chi^2$, where $\chi^2(\alpha_f) =  \sum_i [P_i -
  P(>w_{i,f},\alpha_f)^2]$.

\subsection{Generalized formation redshift and mass accretion history}
The  hierarchical model  predicts that  small haloes  tend to  form at
higher redshift and  then merge together forming the  larger ones. The
halo collapse happens when a  density fluctuation exceeds the critical
value predicted by the spherical  collapse model.  Typically, within a
$\Lambda $CDM Universe  fluctuations in a density field  with a larger
amplitude  collapse earlier.   This can  be rephrased  considering two
identical CDM initial density  fields but with different normalization
parameters $\sigma_8$:  haloes in the Universe  with higher $\sigma_8$
will  collapse  earlier   than  those  in  the   Universe  with  lower
$\sigma_8$.  This also  results in the fact of same  mass haloes being
more  concentrated in  a Universe  with a  larger value  of $\sigma_8$
\citep{maccio08}.  In what  follows we will try to  understand if this
holds also  for different  cosmological scenarios,  i.e.  if  the mass
accretion  history measured  in numerical  simulations of  cDE can  be
reproduced by  the MAH model  built for $\Lambda$CDM only  by suitably
changing the linear power spectrum normalization $\sigma_{8}$.

\begin{figure*}
\begin{center}
\includegraphics[width=7cm]{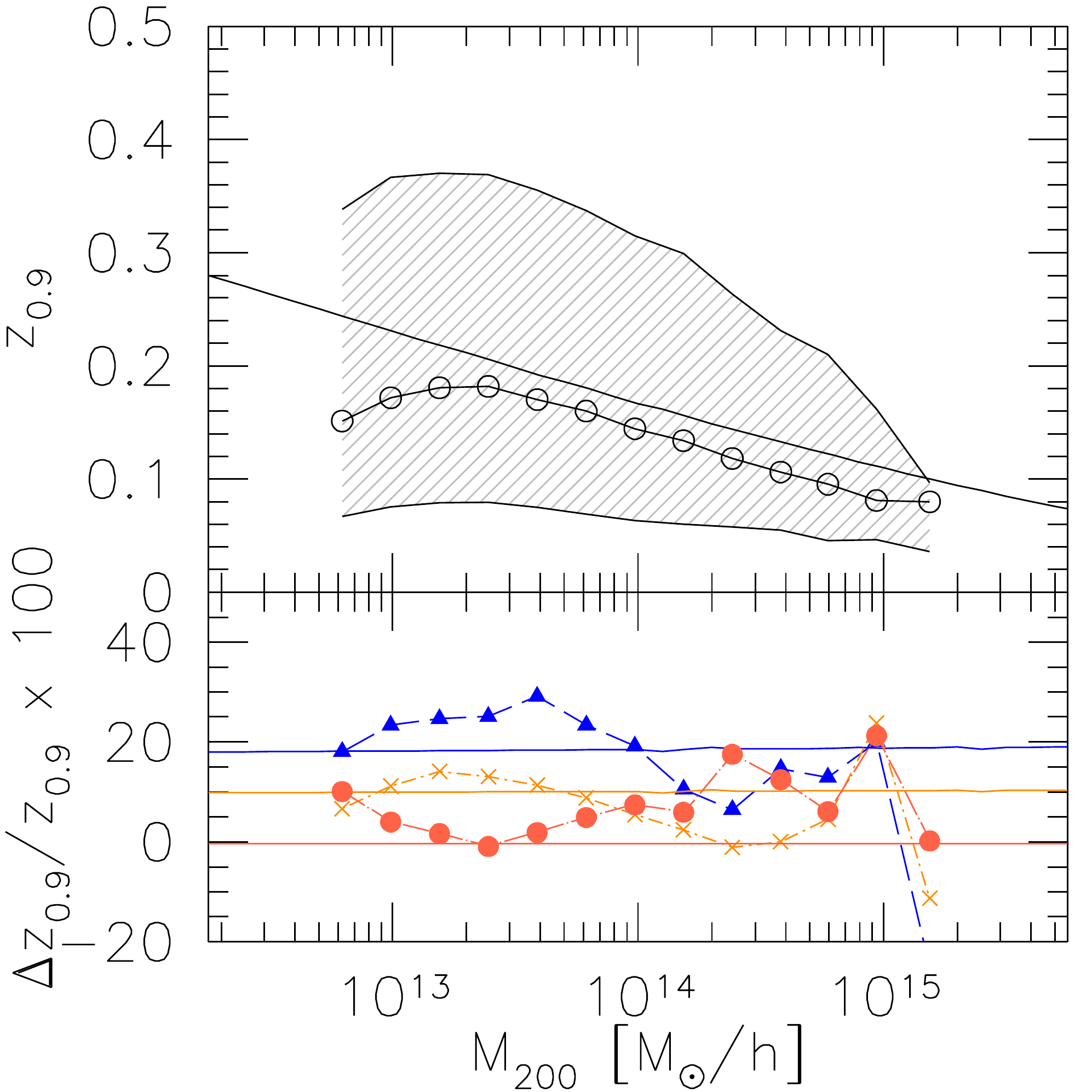} \hspace{1cm}
\includegraphics[width=7cm]{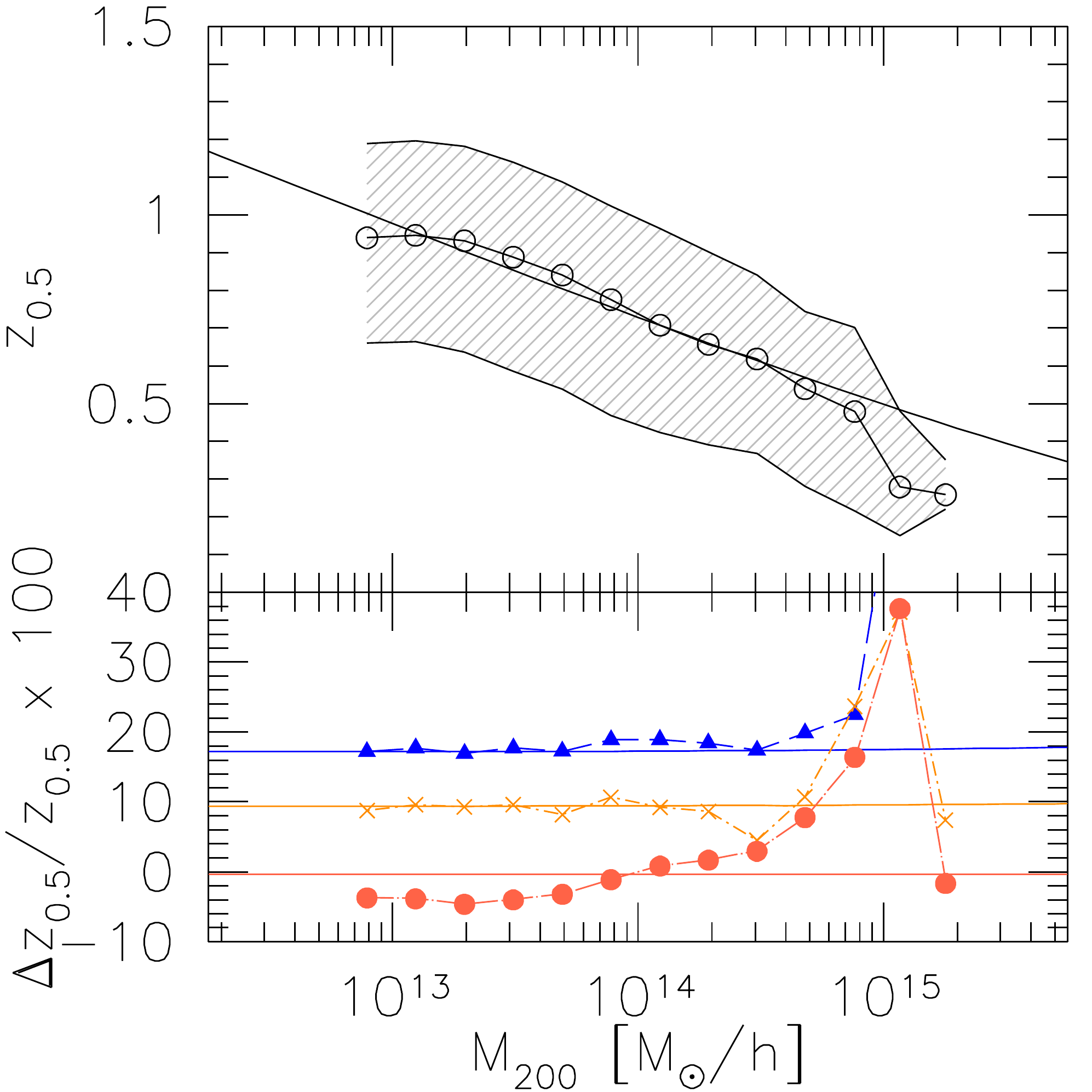} 
\includegraphics[width=7cm]{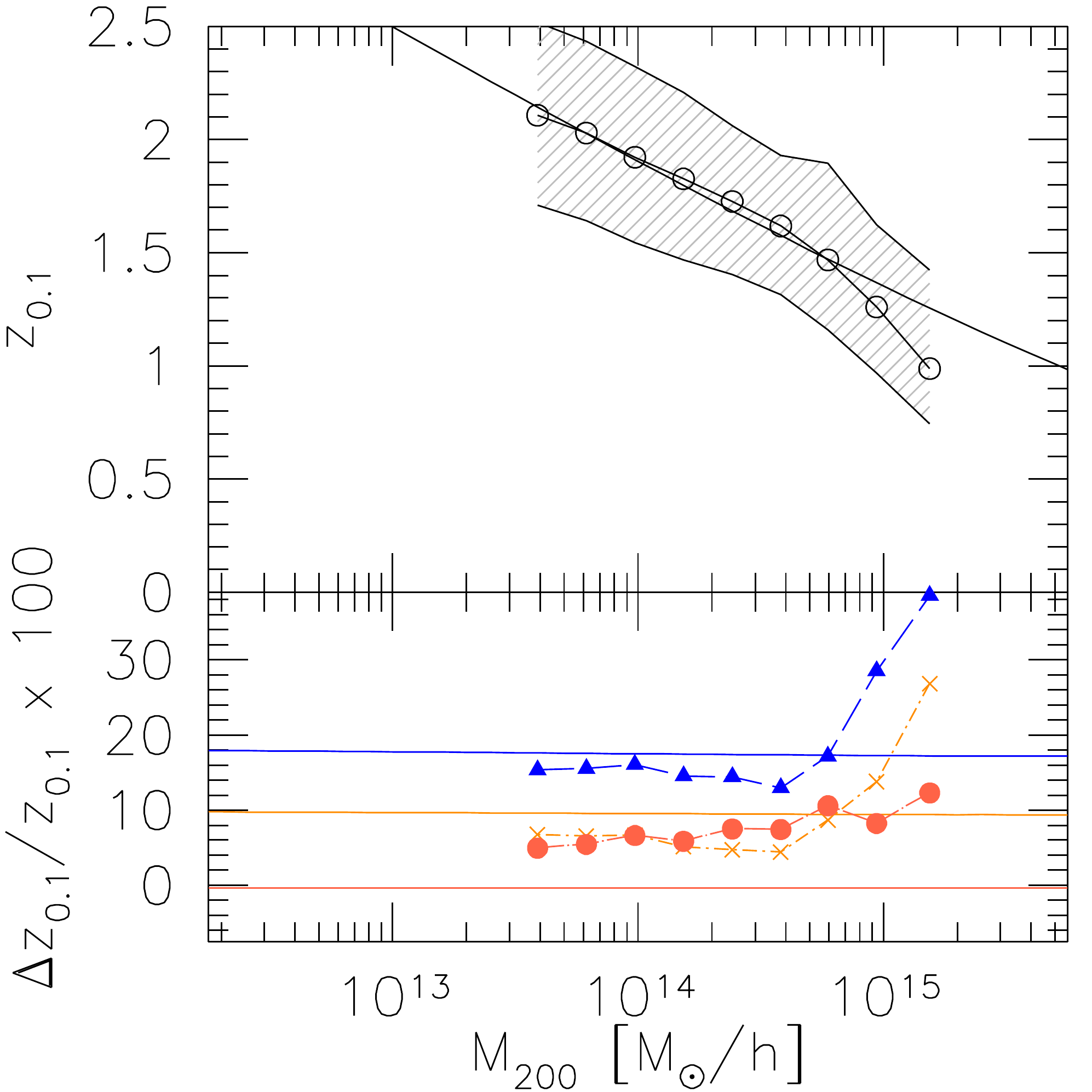} \hspace{1cm}
\includegraphics[width=7cm]{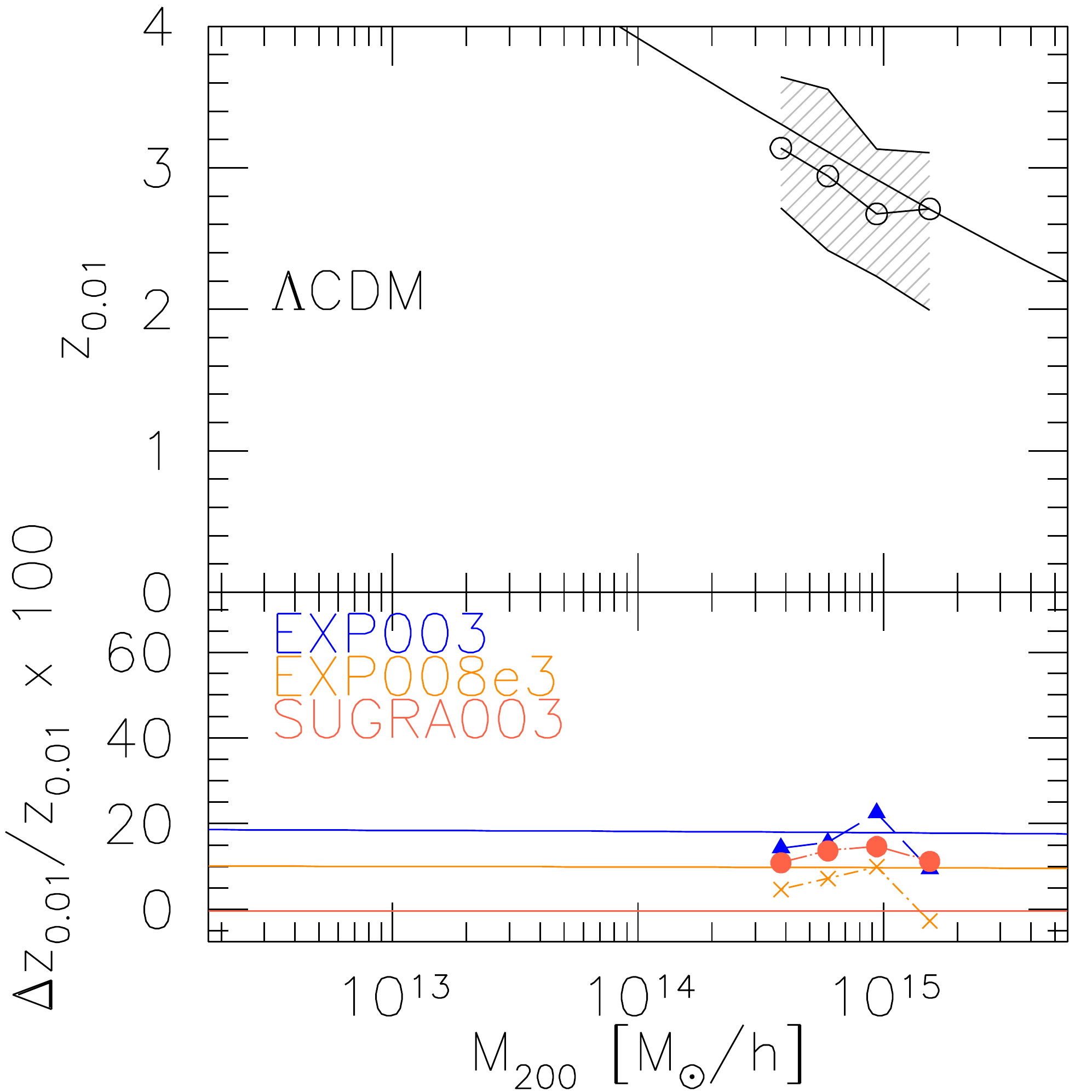}
\caption{Top  panels:  median  formation  redshift-mass  relation  for
  haloes  identified at  $z_0=0$ in  the $\Lambda$CDM  simulation. The
  shaded region  encloses the quartiles  of the distribution  at fixed
  halo mass.  The  solid line represents the  prediction derived using
  the  modified  version  of  the  \citet{giocoli12b}  mass  accretion
  history  model.   Bottom panels:  differences  with  respect to  the
  $\Lambda$CDM model  for the three  cDE models.  The  lines represent
  the   predictions    obtained   from    the   modified    model   by
  \citet{giocoli12b}  by changing  the definition  of $S(M)$  for each
  cosmology only.\label{figformationz}}
\end{center}
\end{figure*}

In the upper part of each  panel of Fig.~\ref{figformationz} we show
the  median  formation  redshift  as  a function  of  the  halo  mass,
considering $f=0.9$, 0.5, 0.1, 0.01  as assembled mass fractions.  The
open   circles  show   the   median  of   the   measurements  in   the
$\mathrm{\Lambda}$CDM simulation  and the  shaded region  encloses the
first and the third quartiles of  the distribution at fixed halo mass.
On   average  a   present-day   cluster  size   halo   with  mass   of
$10^{14}M_{\odot}/h$ assembles  $90\%$ of its mass  at $z=0.2$, $50\%$
at 0.7 and $1\%$ at approximately $z=4$.  The data points in the lower
part of each panel represent the differences of the median measured in
the three  cDE models with  respect to the  $\mathrm{\Lambda}$CDM one;
the  three  curves show  the  model  of the  $z_{f}-M_{200}$  relation
rescaled  with respect  to  the  $\mathrm{\Lambda}$CDM model  computed
using the formalism described in the previous section.

It is important to  underline that in order to build  the MAH model in
the  three  cDE  cosmologies  we  need to  know  the  initial  density
fluctuation field  to compute $S(M)$,  and the redshift  evolution and
the   growth  of   the  perturbations   given  by   $\delta_c(z)$  and
$\Delta_+(z)$.  For the definition of $S(M)$, we use for the three cDE
models  their corresponding  linear power  spectra, that  are obtained
from  the  $\mathrm{\Lambda}$CDM  one   by  renormalizing  the  latter
adopting the  different values  of $\sigma_8$ characterizing  each cDE
model (see Table~\ref{tab:models}).  Since the purpose of this work is
to understand if  the MAH of the cDE simulations  can be obtained from
the $\mathrm{\Lambda}$CDM  by rescaling $\sigma_8$ only,  we adopt for
the definitions  of both $\delta_c(z)$  and $D_{+}$ the  ones obtained
for the  latter model.  From Fig.~\ref{figformationz},  we notice that
only for $f=0.5$ the three cDE results are quite in agreement with the
MAH model where the power spectrum normalization changes.  Among them,
since   the    SUGRA003   model    has   the   same    $\sigma_8$   of
$\mathrm{\Lambda}$CDM, we  would expect  to find  a formation-redshift
relation quite  similar to the  latter.  However, this is  clearly not
the case due to the markedly  different growth of SUGRA003 as compared
to  $\Lambda $CDM.   For $f=0.9$  and $f=0.5$  haloes in  the SUGRA003
model   form   typically   at   the    same   redshift   as   in   the
$\mathrm{\Lambda}$CDM run, while for $f=0.1$ and $0.01$ the difference
appears of  the order of  $10\%$. We underline  also that for  the two
models EXP003 and EXP008e3 the halo formation redshifts are higher for
any $f$ than those measured in $\mathrm{\Lambda}$CDM for the same mass
haloes at $z=0$.

\begin{figure*}
\begin{center}
\includegraphics[width=5.8cm]{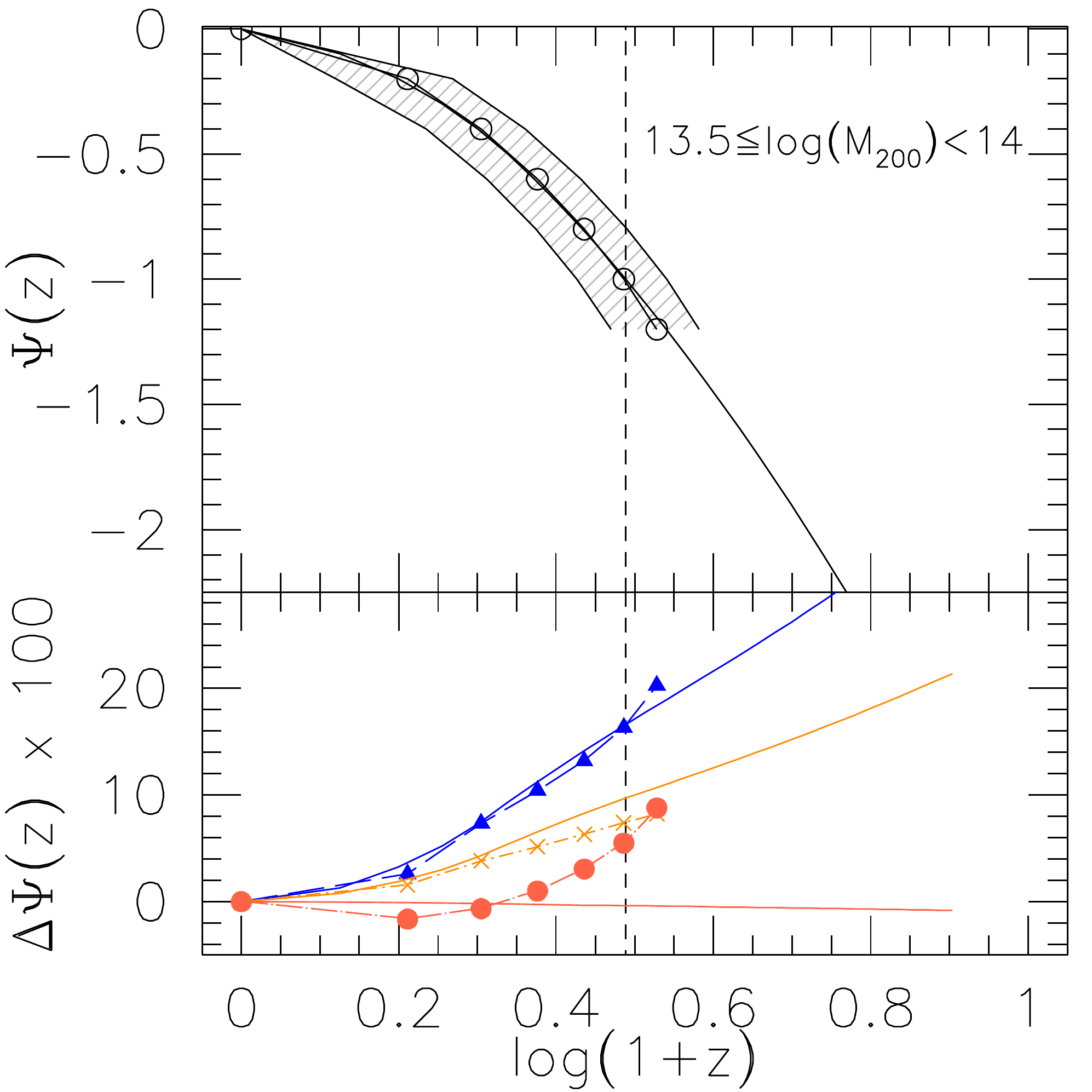}
\includegraphics[width=5.8cm]{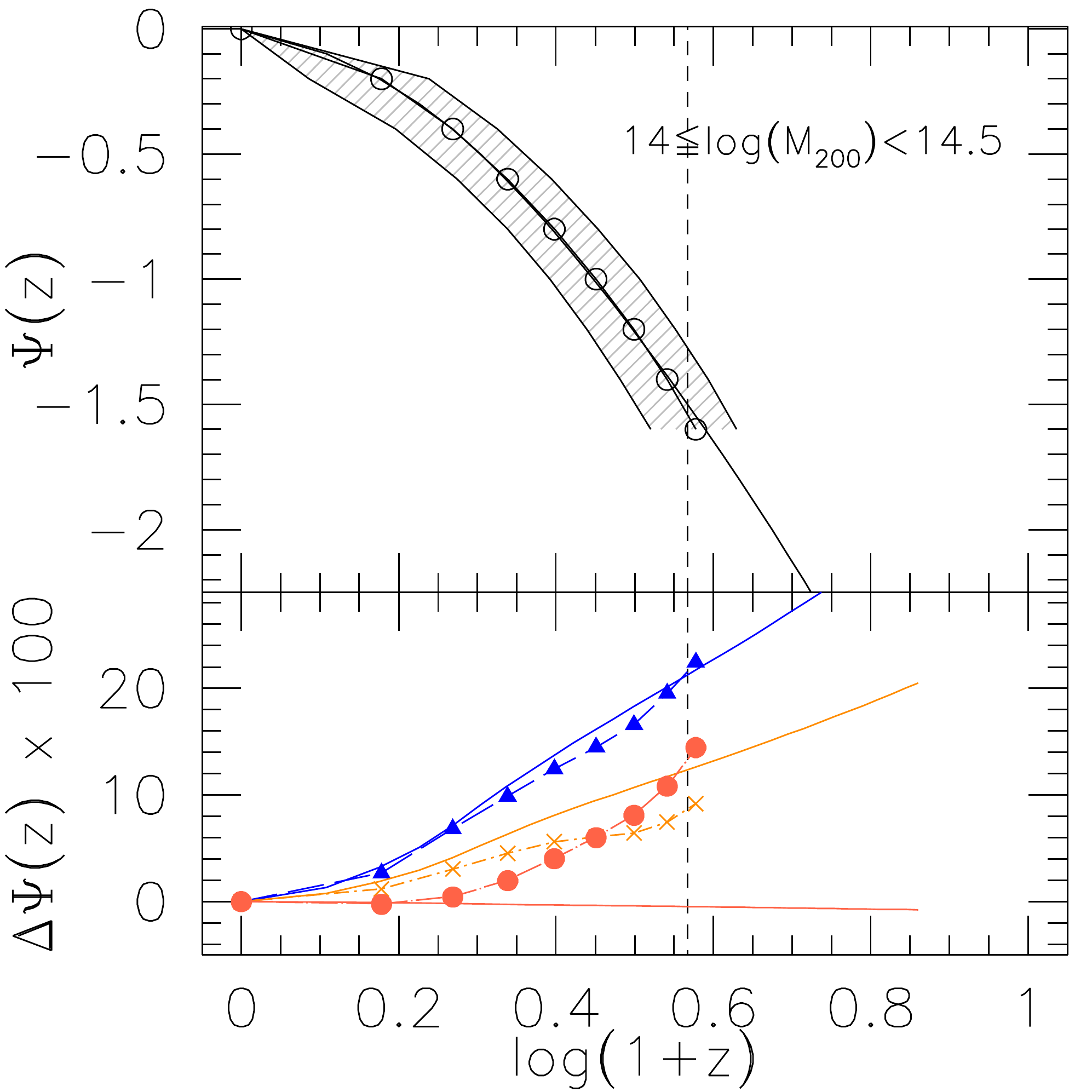}
\includegraphics[width=5.8cm]{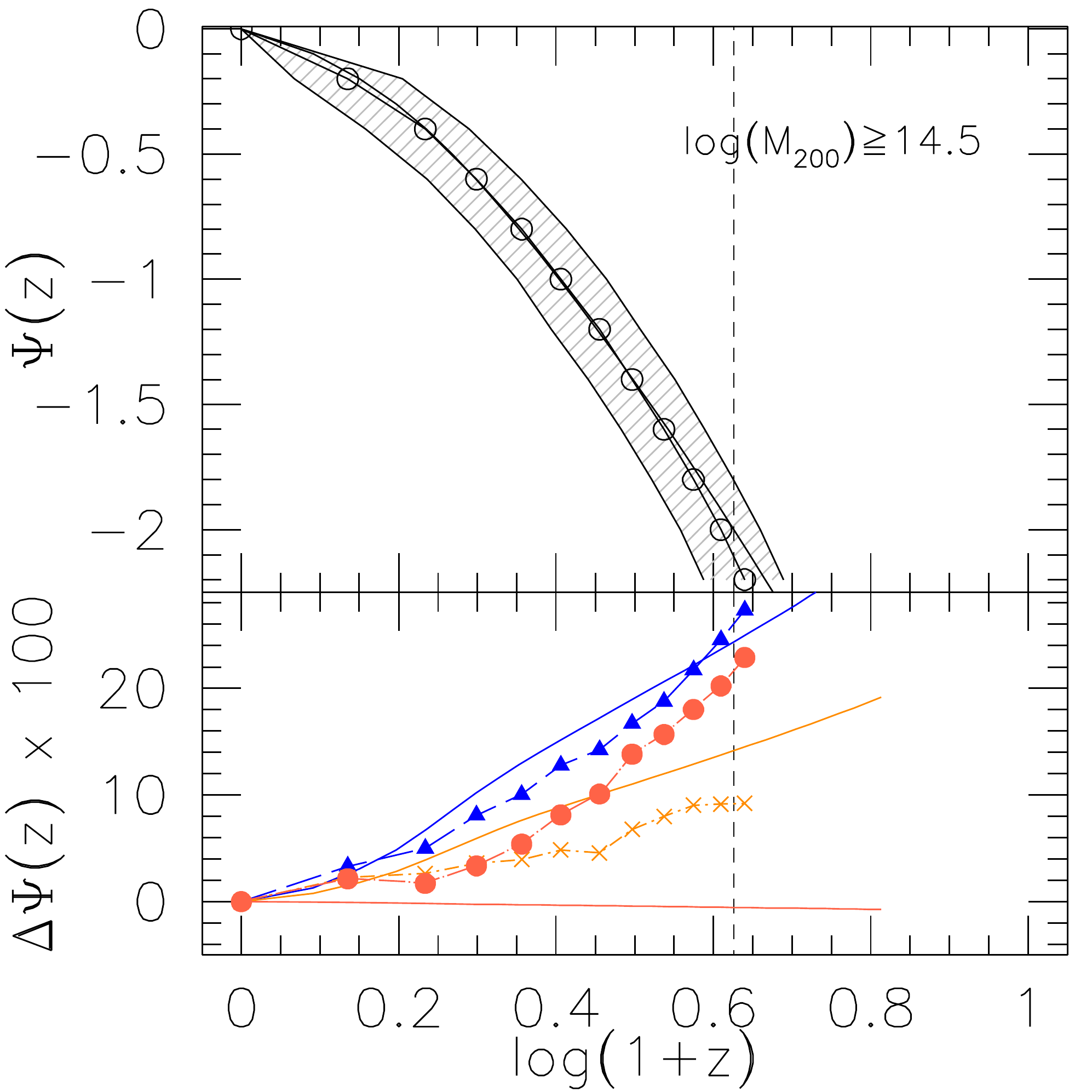}
\caption{Median mass accretion history of main halo progenitors. In
  the top part of each panel we show the results from the
  $\mathrm{\Lambda}$CDM simulation for three different initial mass
  bins, where the shaded region encloses the quartiles of the
  distribution at fixed halo mass fraction assembled.  In each panel,
  the solid curve represents the prediction from
  eq.~(\ref{eqmahmodz}).  The bottom part of each panel shows the
  residuals with respect to the $\mathrm{\Lambda}$CDM simulation
  measurements of the three cDE models together with the residuals for
  the theoretical predictions changing $S(M)$ in the mass accretion
  history model.  Point style as in
  Fig.~\ref{figformationz}. \label{figMAH}}
\end{center}
\end{figure*}

To better understand  how on average haloes assemble as  a function of
redshift, we  show in  Fig.~\ref{figMAH}  the median  mass accretion
history for three  different mass bins at $z=0$ by  defining $\Psi$ as
the logarithm of the final assembled mass fraction:
\begin{equation}
\Psi(z) = \log \left[ \dfrac{M_{200}(z)}{M_{200}(0)}\right] =
\log(f)\,.
\end{equation}
In the top  panels we show the measurements for  three different final
halo  mass  bins at  the  present  time in  the  $\mathrm{\Lambda}$CDM
simulation.   The  shaded region  encloses  the  first and  the  third
quartiles of the distribution at  fixed halo mass fraction.  The solid
curve represents the model  built using equation~(\ref{eqmahmodz}) for
different values of the assembled halo mass fraction $0<f<1$.

The model predicts for the $\mathrm{\Lambda}$CDM  run a MAH that is in
agreement  within   a  few  percent  with   the  numerical  simulation
measurements  both down  to very  small values  of the  assembled mass
fraction and up  to high redshifts.  In the bottom  panels we show the
percentage  difference of  the MAH  measured in  EXP003, EXP008e3  and
SUGRA003 with  respect to  the one  in $\mathrm{\Lambda}$CDM.   In the
figure,   the   different   data   points   are   the   same   as   in
Fig.~\ref{figformationz}.  The   three  solid  curves   represent  the
difference of MAH models changing the linear power spectrum definition
with respect  to the $\mathrm{\Lambda}$CDM model.   Since the SUGRA003
model has almost  the same linear power spectrum  normalization of the
$\mathrm{\Lambda}$CDM, the MAH  model predicts the same  growth of the
latter, in contrast with what is measured in the numerical simulation.
For EXP003 the halo growth is  quite well captured changing the linear
power spectrum  normalization in  our model -  and is  therefore fully
degenerate with $\sigma_8$.  However, for EXP008e3 the model fails for
large haloes and high redshift.

\section{The Halo Structural Properties}
\label{secHaloProp}

\subsection{Halo concentrations}
 
One of the  most important result obtained  through N-body simulations
of  structure  formation  is  that the  CDM  density  distribution  in
collapsed haloes tends  to follow a universal profile.  Both for small
and large  mass haloes the  density profile  is well described  by the
Navarro-Frenk-White \citep[NFW,][]{navarro97} relation that reads as:
\begin{equation}
\rho(x) = \dfrac{\rho_s}{x (1 + x)^2}\,,
\end{equation}
where  $x\equiv  r/r_s$, $r_s  \equiv  R_{200}/c_{200}$  is the  scale
radius where the  logarithmic slope of the  density profile approaches
$-2$, $\rho_s$  is the density  enclosed within the scale  radius, and
$c_{200}$  is   the  halo  concentration  parameter.    Denoting  with
$R_{200}$ the radius enclosing $M_{200}$ we can write:
\begin{equation}
\rho_s = \dfrac{M_{200}}{4 \pi r_s^3} \left[ \ln(1+c_{200}) - 
\dfrac{c_{200}}{1+c_{200}}\right]^{-1}\,.
\end{equation}

\begin{figure}
\begin{center}
\includegraphics[width=\hsize]{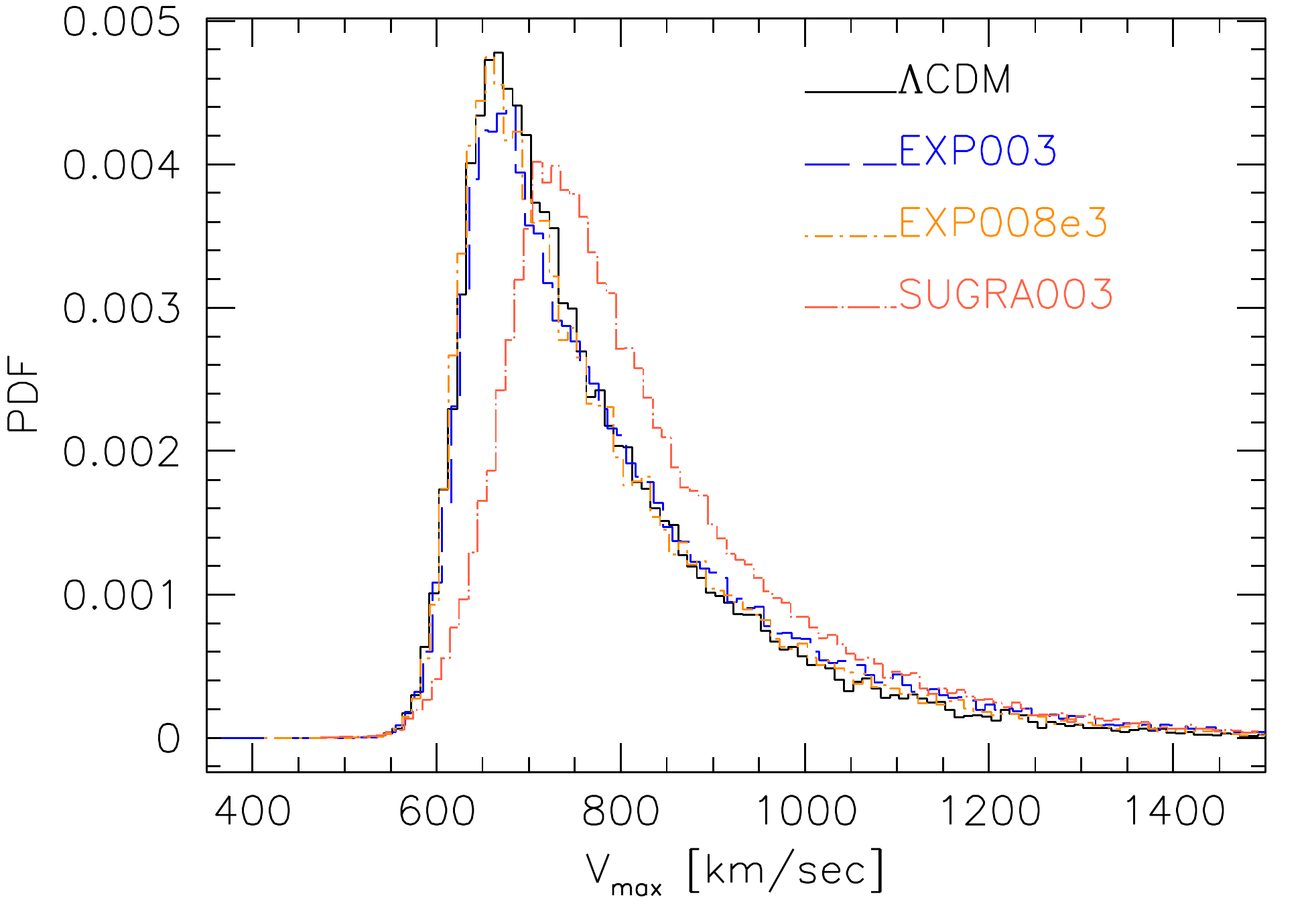}
\includegraphics[width=\hsize]{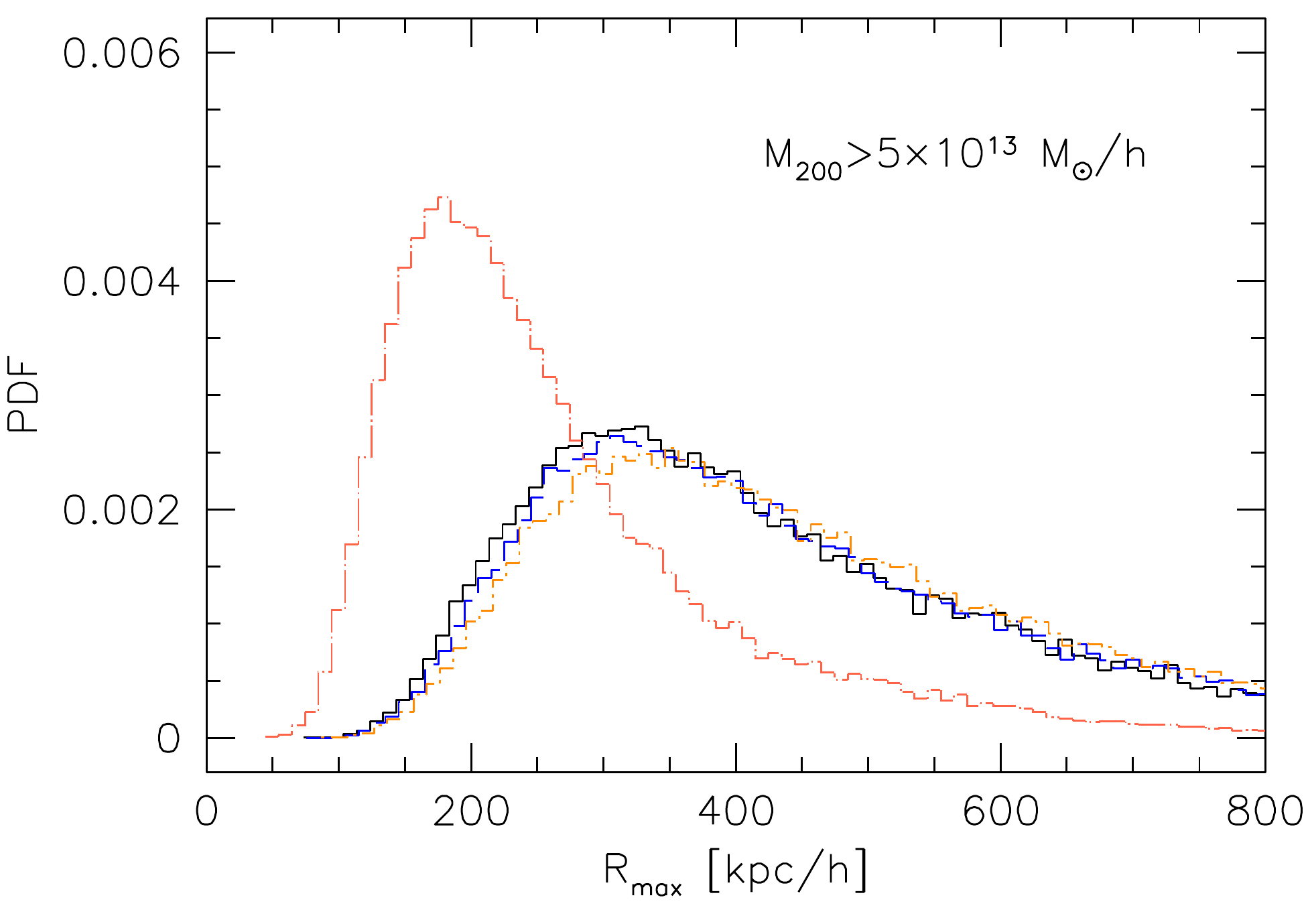}
\caption{\label{figmaxdist} Distribution function  of maximum circular
  velocity (top  panel) and  of the  radius at  which this  is reached
  (bottom panel)  for all haloes at  $z=0$ with masses larger  than $5
  \times 10^{13}M_{\odot}/h$ in the four cosmological simulations. }
\end{center}
\end{figure}

To characterize the halo concentration in the numerical simulations we
use   the   same   approach   adopted   by   \citet{springel08b}   and
\citet{cui12}: defining $V_{max}$ as  the maximum circular velocity of
a halo and $r_{max}$ as the  radius at which this velocity is attained
so we can write:
\begin{eqnarray}
\dfrac{\rho_s}{\rho_{crit}} &=& \dfrac{200}{3}
\dfrac{c_{200}^3}{\ln(1+c_{200}) - c_{200}/(1+c_{200})} \nonumber
\\ &=& 14.426 \left( \dfrac{V_{max}}{H(z) \, r_{max}}\right)^2\,,
\end{eqnarray}
where   $\rho_{crit}$  represents   the   critical   density  of   the
Universe. In Fig.~\ref{figmaxdist} we show the distribution function
of the  maximum circular  velocity and  of the radius  at which  it is
attained for all haloes that at the present time are more massive than
$5 \times 10^{13}M_{\odot}/h$.  Since to faithfully determine the halo
concentration   a   minimum   number    of   particles   is   required
\citep{neto07}, for this analysis we consider only haloes more massive
than $5 \times 10^{13} M_{\odot}/h$; this lower limit ensures that the
haloes are  resolved with  at least $850$  particles.  We  notice that
while the haloes  in EXP003 and EXP008e3 have values  of $V_{max}$ and
$R_{max}$ that are not much different from the corresponding haloes in
the $\mathrm{\Lambda}$CDM, the haloes in SUGRA003 are more compact and
typically have a higher maximum  circular velocity that is attained at
a smaller radius  than in $\mathrm{\Lambda}$CDM; as we  will see later
on, this corresponds to more  concentrated haloes in SUGRA003 at $z=0$
than in the other three cosmological models.

In the context  of the CDM hierarchical  structure formation scenario,
halo concentrations are  thought to be reminiscent of  the cosmic mean
density at the time of  collapse, thereby resulting in smaller objects
having on average higher concentrations due to their earlier formation
epoch. Consistently with  this picture, clusters of  galaxies have the
lowest concentrations --  typically of the order of $\sim  4$ -- being
the last collapsed structures in  the Universe.  At fixed redshift and
halo  mass  the concentration  tends  to  be  larger for  haloes  that
assemble their mass earlier; this also corresponds to haloes that are,
on average,  more relaxed.   \citet{neto07}, \citet{maccio07,maccio08}
and \citet{deboni12} have shown that including unrelaxed haloes in the
sample results  in a $c-M$ relation  with a lower normalization  and a
larger scatter.

Several studies  based on the  analysis of numerical  simulations have
shown that halo  structural properties are mainly related  to the halo
mass    accretion   history,    and    so    to   their    environment
\citep[][]{gao04,sheth04a,gao08}.   Not only  do  less massive  haloes
possess  a higher  concentration  but typically  also  a smaller  mass
fraction  in substructures  \citep{delucia04,vandenbosch05,giocoli10}.
In   particular,   the   $c-M$   models   by   \citet{navarro97}   and
\citet{bullock01a} rely  on the idea  that the central density  of the
haloes reflects  the mean density of  the Universe at a  time when the
central  region   of  the  halo   was  accreting  matter  at   a  high
rate. \citet{zhao09} define  this epoch when the  main halo progenitor
first containing $4\%$ of the final mass.

In  order to  understand how  halo concentration  correlates with  the
accretion history,  in what follows  we study the  correlation between
$c_{200}$ and two conventional markers of the halo MAH: $z_{0.5}$ i.e.
the redshift at  which the main halo progenitor assembles  half of its
final mass  (we will  denote with  $t_{0.5}$ the  corresponding cosmic
time)  and $z_{0.04}$  (at  which corresponds  $t_{0.04}$), i.e.   the
redshift at which it assembles $4\%$ of its mass.

\begin{figure*}
\begin{center}
\includegraphics[width=7cm]{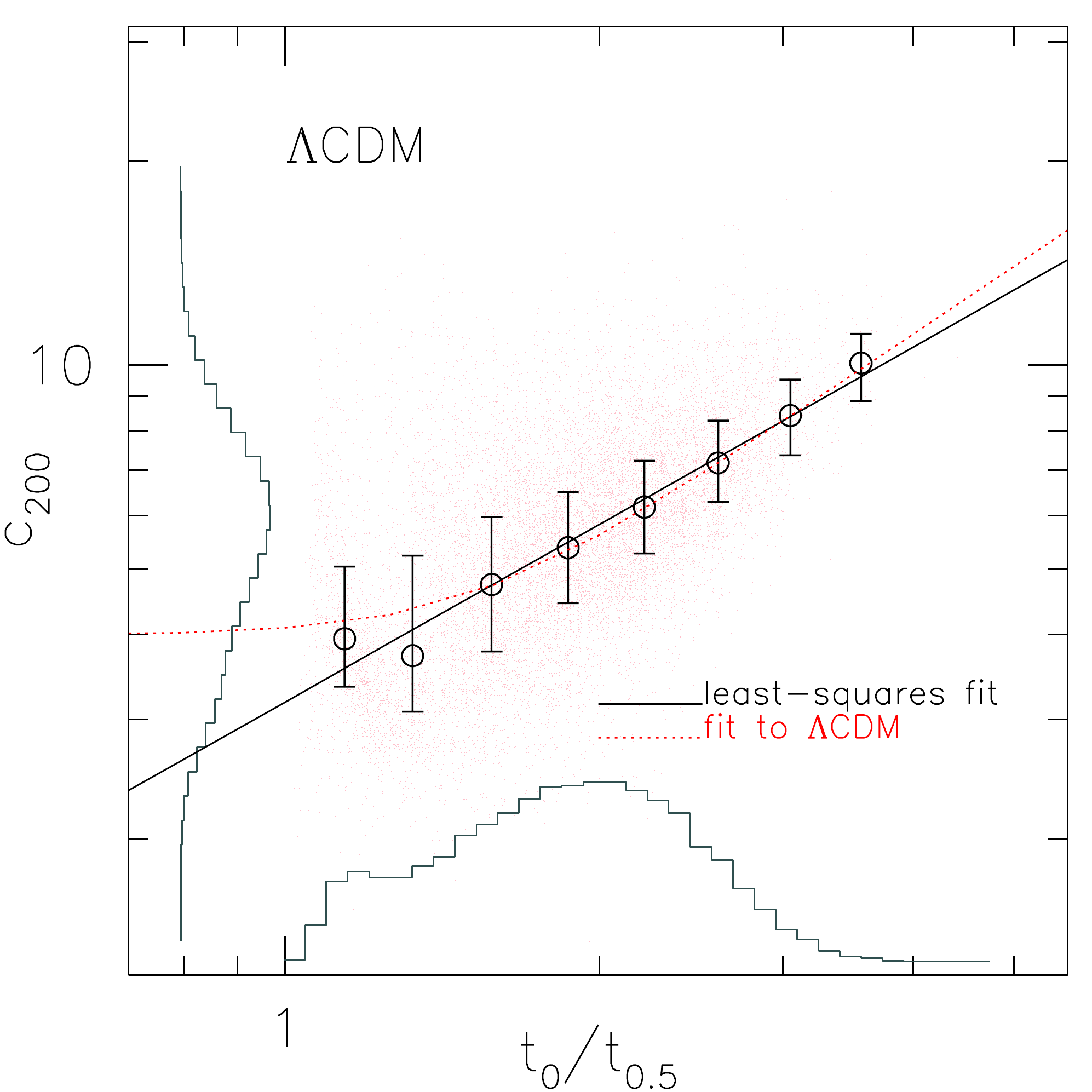} \hspace{1cm}
\includegraphics[width=7cm]{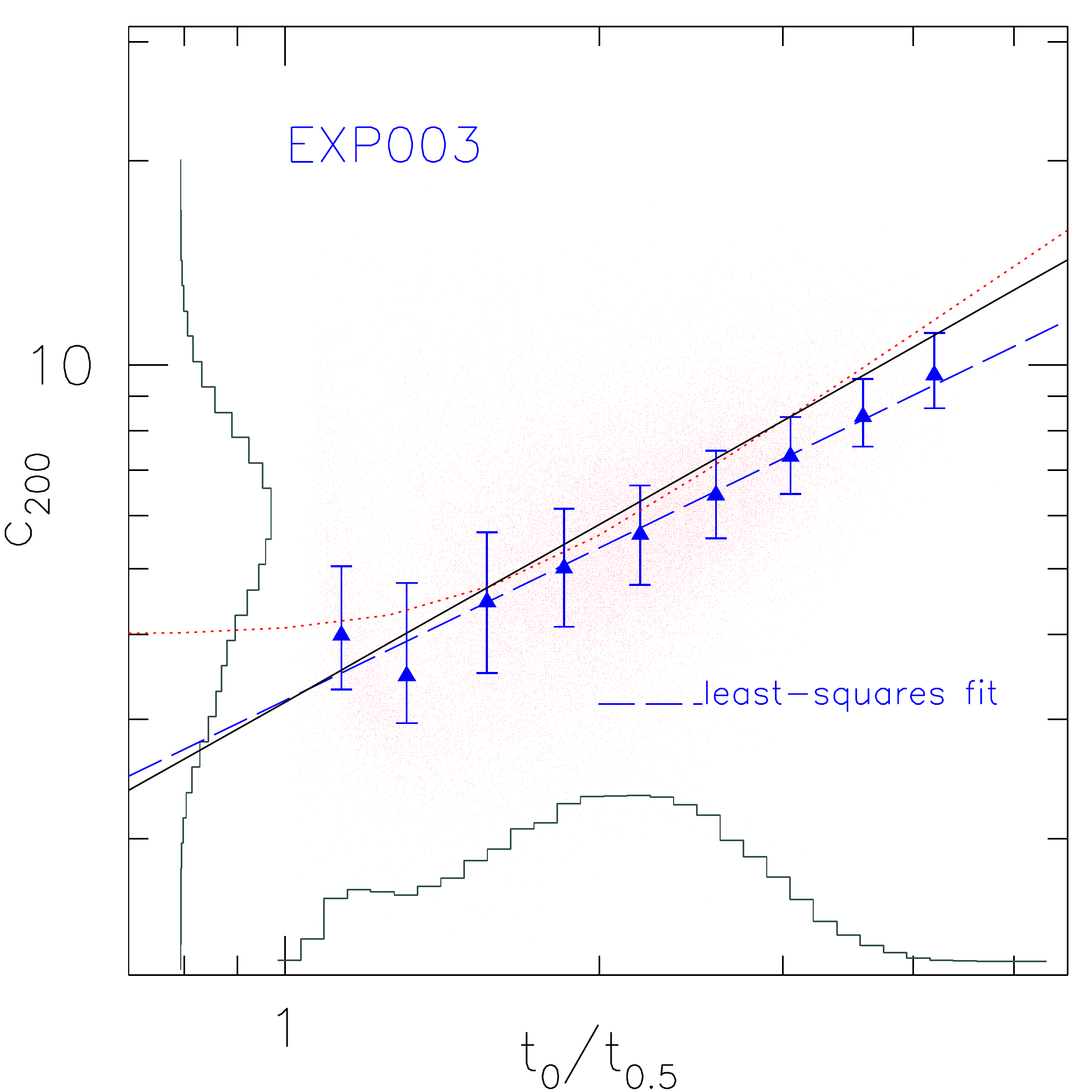}
\includegraphics[width=7cm]{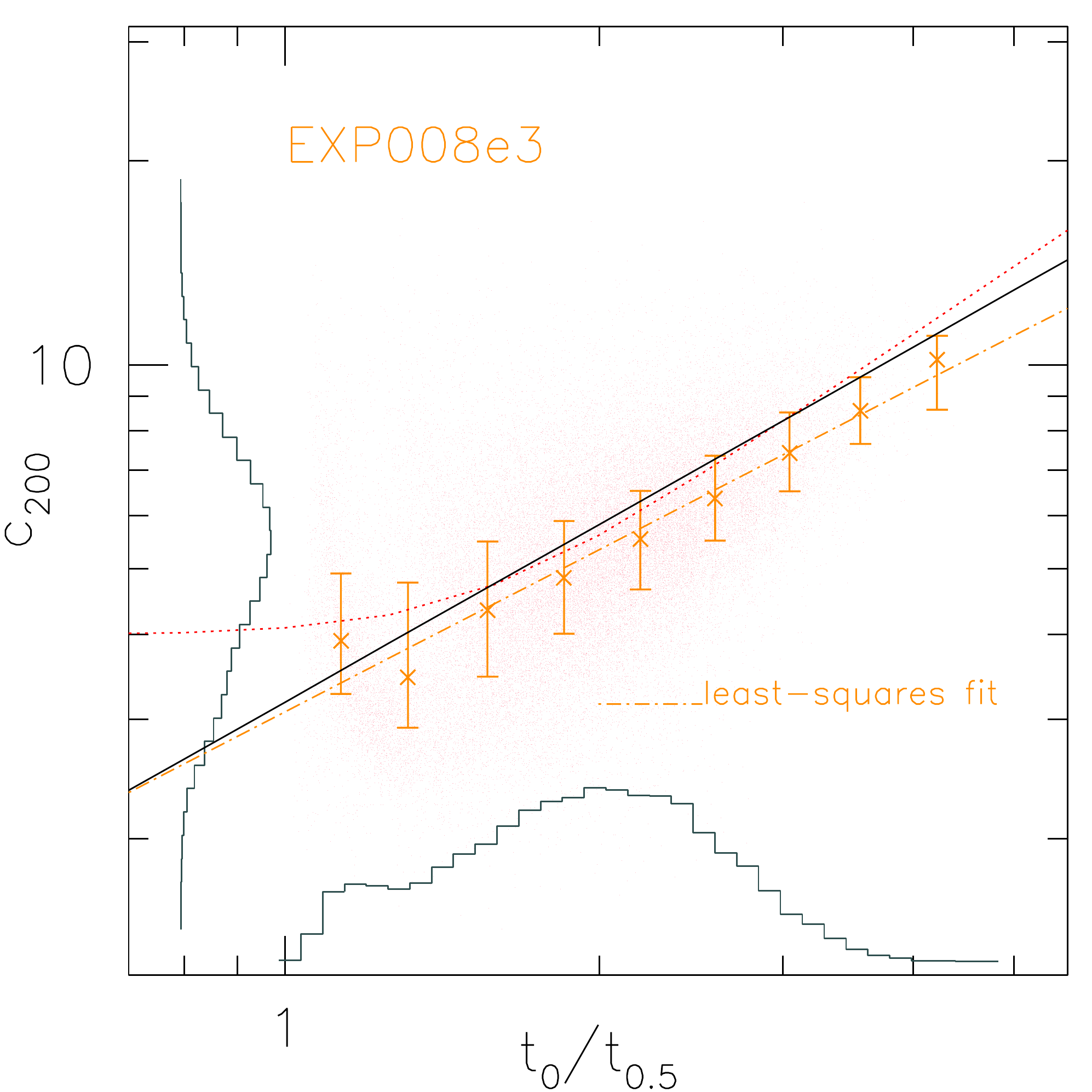} \hspace{1cm}
\includegraphics[width=7cm]{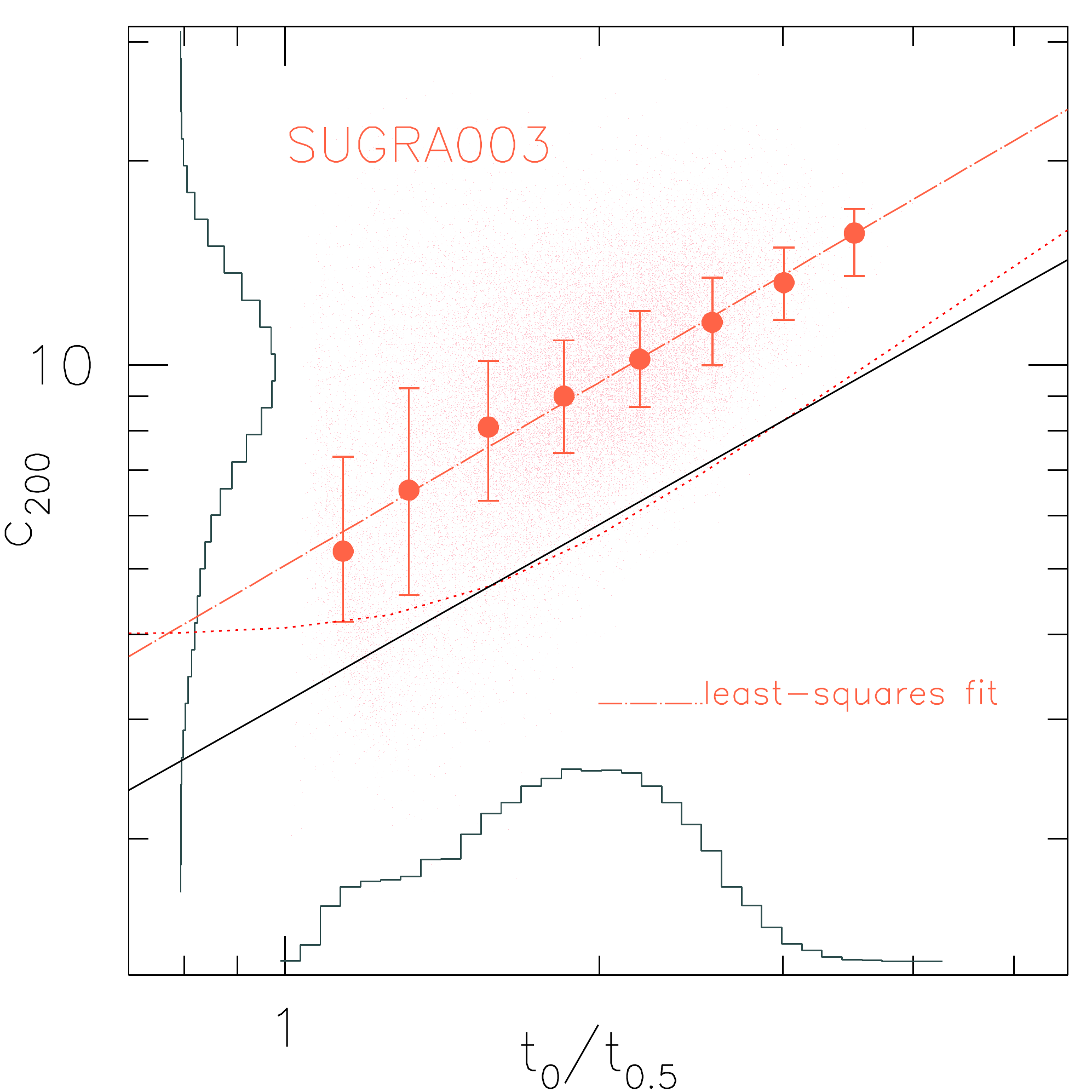} 
\caption{Correlation between the concentration  and the formation time
  when  the main  halo  progenitor  assembles $50\%$  of  the mass  at
  $z_0=0$. The four panels refer to the different cosmological models.
  In each  panel the points with  the error bars show  the median with
  the two quartiles.  Solid (black) and dotted lines in all panels are
  the same  and represent the  least-square fit  to the haloes  in the
  $\Lambda$CDM  simulation  and equation~(\ref{eqct05})  respectively.
  The different line  types in the panels referring to  the cDE models
  represent  the   corresponding  least-square   fit  to   their  data
  points.  \label{figcmt05}}
\end{center}
\end{figure*}

In  Fig.~\ref{figcmt05}  we show  the correlation  between the  halo
concentration  and $t_0/t_{0.5}$  where $t_0$  represents the  time at
which the halo is considered -- in our case $z_0=0$.  In each panel we
show  the measurements  for all  haloes  more massive  than $5  \times
10^{13}M_{\odot}/h$  in the  $\mathrm{\Lambda}$CDM and  the three  cDE
models.   The data  points  with error  bars show  the  median of  the
correlation   and  the   quartiles  of   the  distribution   at  fixed
$t_0/t_{0.5}$.  In the  top-left panel the solid black  line shows the
least-square fit  to the  $\mathrm{\Lambda}$CDM measurements,  and the
dotted red curve represents  the following best-fit power law-relation
obtained by minimizing the scatter:
\begin{equation}
c_{200} = 4 \left[ 1+ \left( \dfrac{t_0}{1.5 t_{0.5}}\right)^{5.2}\right]^{1/5}\,.
\label{eqct05}
\end{equation}
Equation~(\ref{eqct05})   is   similar   to  the   one   proposed   by
\citet{zhao09}.  These two relations are also overplotted in the other
panels.  In each panel the grey histograms on the $x$ and the $y$ axis
represent   the   distributions   of  $t_0/t_{0.5}$   and   $c_{200}$,
respectively.  The  different line types  in the EXP003,  EXP008e3 and
SUGRA  panels   show  the  least-squares  fit   to  the  corresponding
measurements.  In  Table \ref{tabct05} we summarize  these results and
give  an estimate  of the  rms of  the data  points for  the different
numerical simulation results defined as:
\begin{equation}
\mathrm{rms} = \sqrt{\sum_{i=1}^N \dfrac{\left( \log(c_i) - \log(c_{model})\right)^2}{N}}\,.
\end{equation}

\begin{table*}
\caption{ \label{tabct05}Rms of the data with respect to the models of
  Fig.~\ref{figcmt05}.}
\begin{center}
\begin{tabular}{|l|c|c|c|r|}
\hline
\hline
model & eq.~(\ref{eqct05})  & $\mathrm{\Lambda}$CDM
least-squares & least-squares & least-square values \\ 
$ $ & $ $ & $ $ & $ $ & slope, zero point   \\ \hline
$\mathrm{\Lambda}$CDM &  0.128 & 0.127 & - & 0.870, 0.503 \;\;\;\;\;\;\\
EXP003 & 0.132 & 0.134 & 0.128 & 0.748, 0.506 \;\;\;\;\;\;\\
EXP008e3 & 0.133 & 0.135 & 0.128 & 0.749, 0.488 \;\;\;\;\;\;\\
SUGRA003 & 0.255 & 0.254 & 0.135 & 0.897, 0.704 \;\;\;\;\;\;\\
\hline
\hline
\end{tabular}
\end{center}
\end{table*}

\begin{figure*}
\begin{center}
\includegraphics[width=7cm]{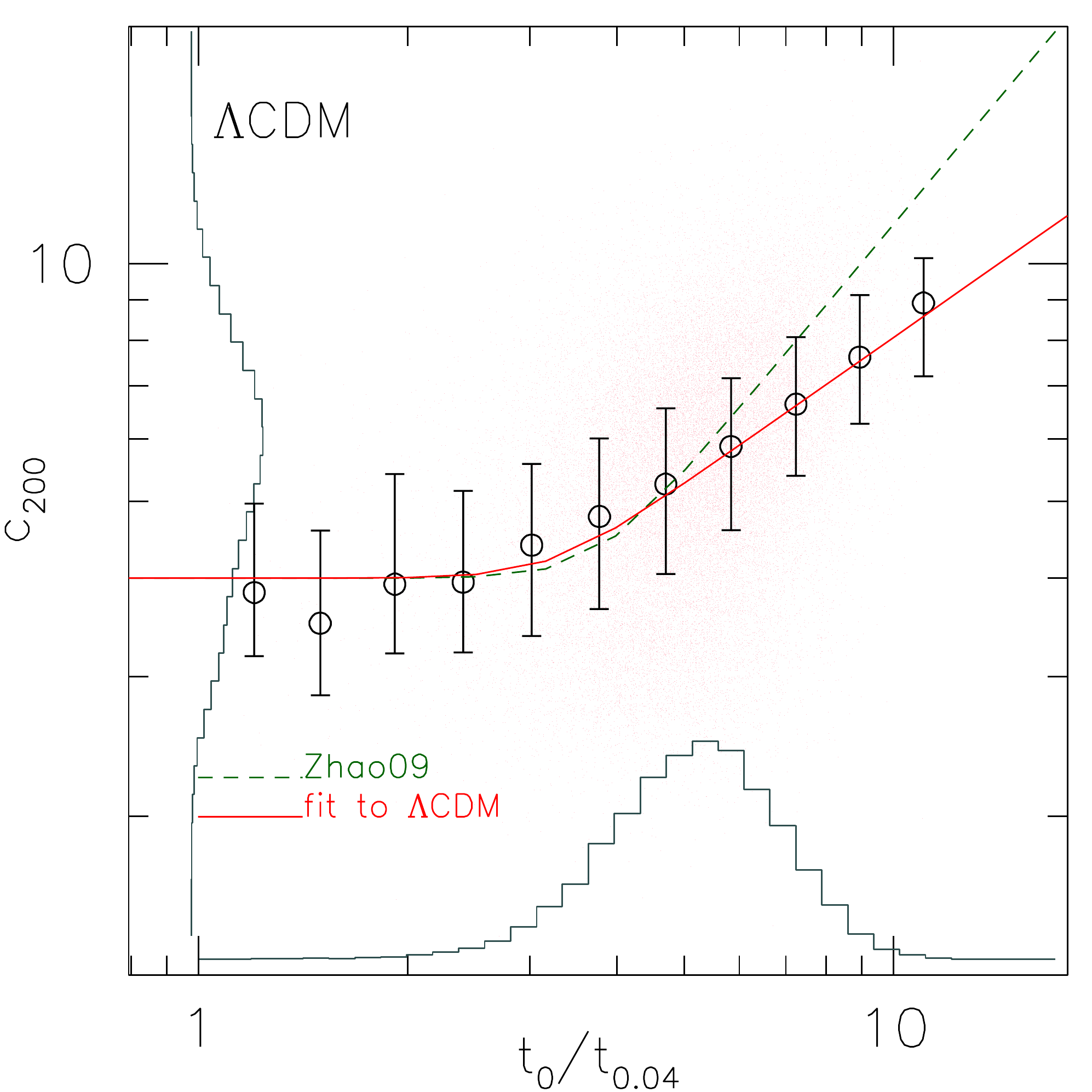} \hspace{1cm}
\includegraphics[width=7cm]{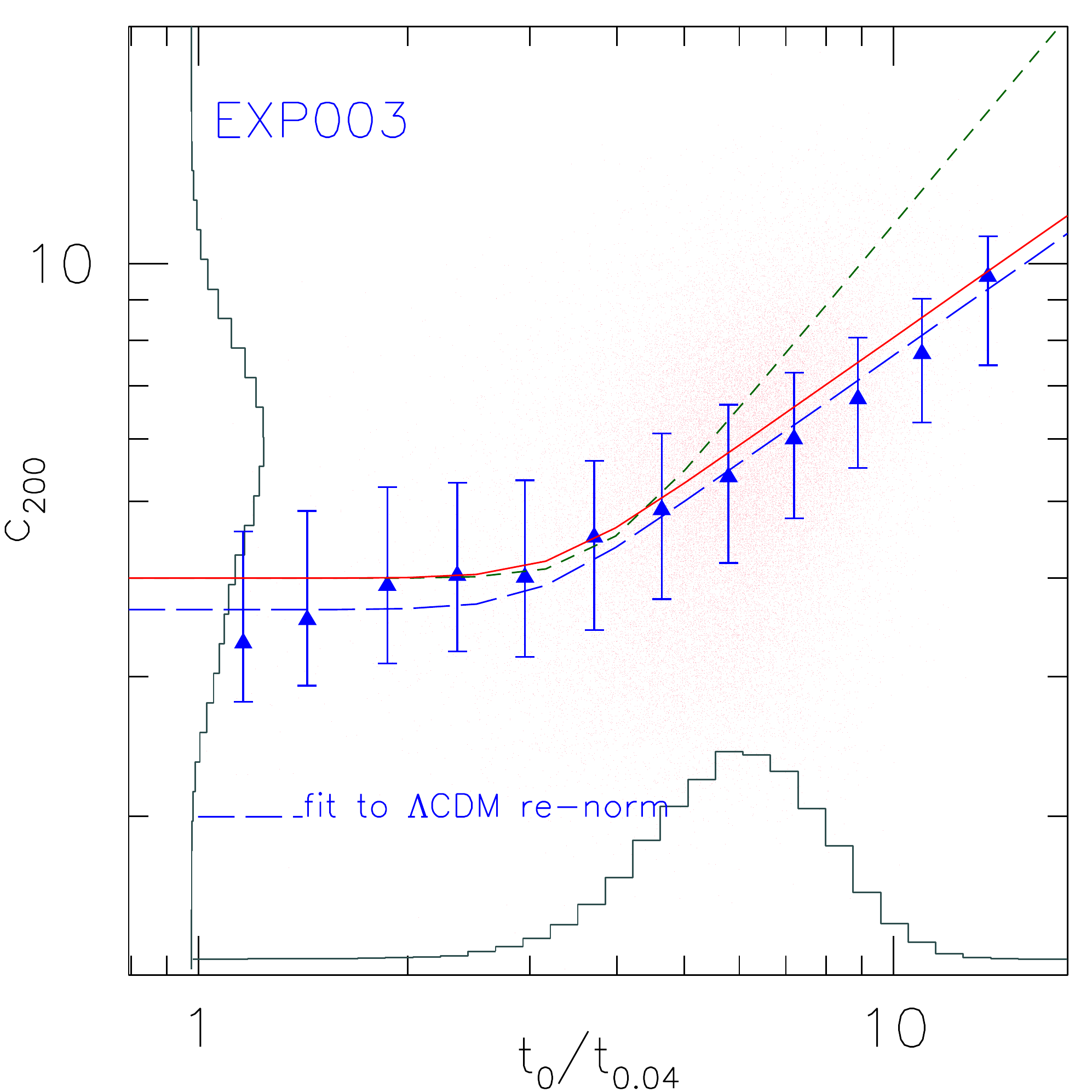}
\includegraphics[width=7cm]{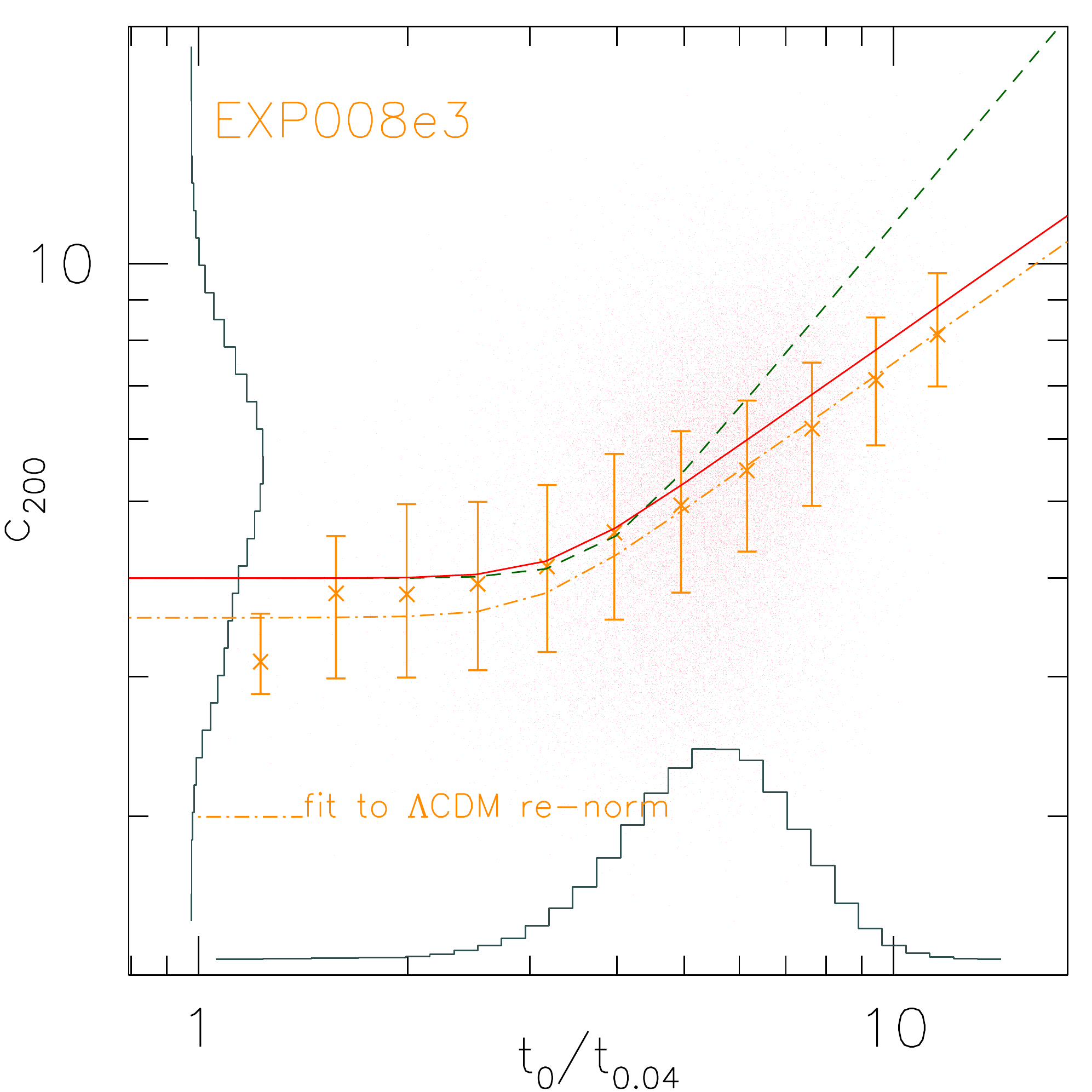} \hspace{1cm}
\includegraphics[width=7cm]{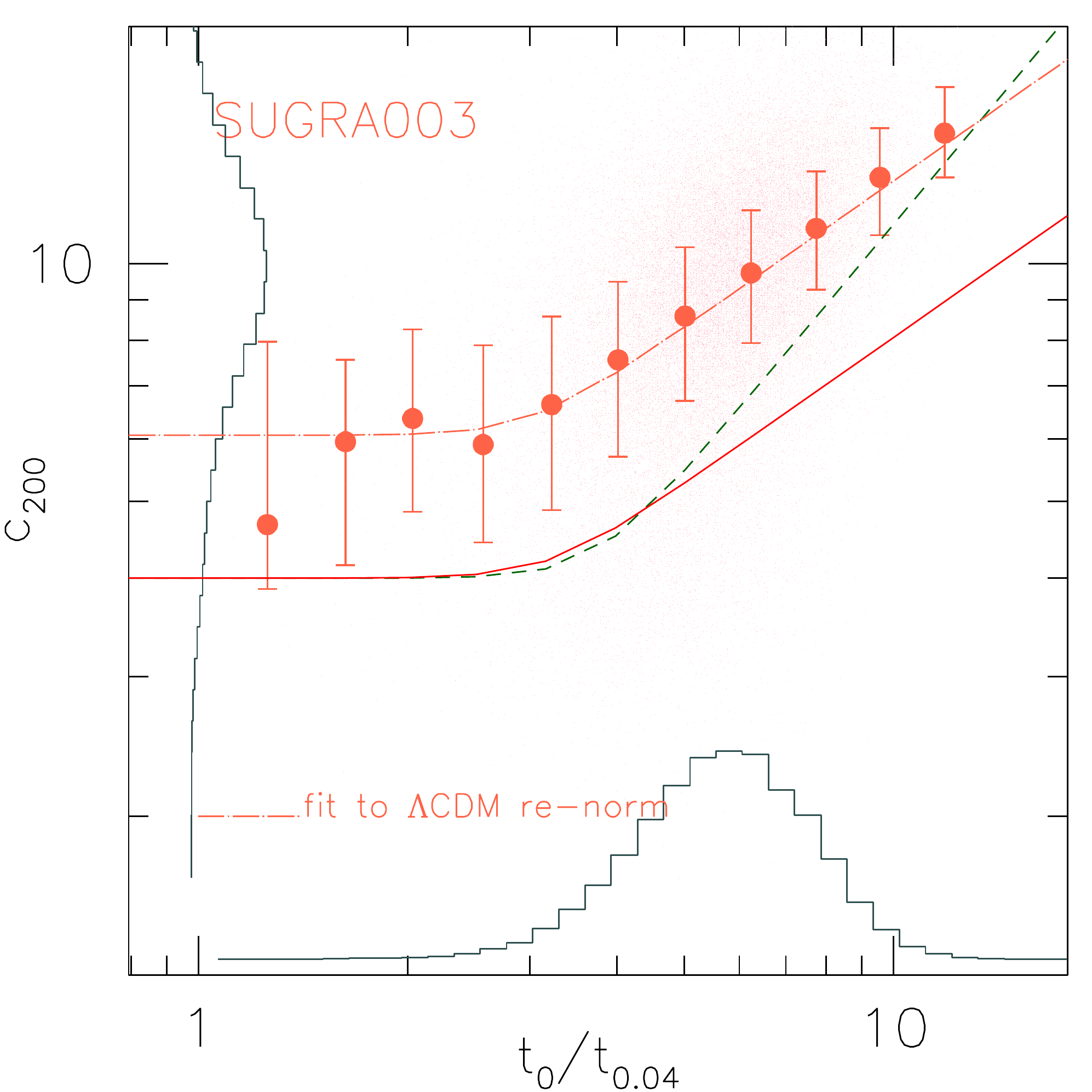}
\caption{Correlation between the concentration and the formation time
  when the main halo progenitor assembles $4\%$ of the mass at
  $z_0=0$.  Dashed and solid curves in all panels show
  equations~(\ref{eqzhao09}) and (\ref{eqct004}). In the three cDE
  models the different line type curves show equation~(\ref{eqct004})
  renormalized to best-fitting the corresponding data
  points. \label{figcmt004}}
\end{center}
\end{figure*}

\citet{zhao09} have  shown that  the halo  concentration has  a strong
correlation with the time $t_{0.04}$ at which the main halo progenitor
assembles \textbf{for the first time} $4\%$ of its mass, in the idea that
typically haloes  acquire a  concentration of  $4$ when  they assemble
$4\%$ of their  mass.  From their measurements, the  best-fitting relation
between the concentration and $t_{0.04}$ reads as:
\begin{equation}
c_{vir} = 4  \left[ 1+ \left( \dfrac{t_0}{3.75 t_{0.04}} \right)^{8.4} \right]^{1/8}\,,
\label{eqzhao09}
\end{equation}
where $c_{vir}$  represents the ratio  between the scale  radius $r_s$
and the virial radius $R_{vir}$.  We recall that \citet{zhao09} define
$R_{vir}$  such   that  it  encloses  an   overdensity  $\Delta_{vir}$
according to the spherical collapse  model, with respect to which they
also define the  virial mass $M_{vir}$.  In  Fig.~\ref{figcmt004} we
show   the   median  of   the   correlation   between  $c_{200}$   and
$t_0/t_{0.04}$  for the  measurements in  the four  N-body simulations
used    in     this    work.     The    short-dashed     lines    show
equation~(\ref{eqzhao09}), that  we recall is valid  for the $c_{vir}$
and  $M_{vir}$ definitions.   In the  top-left panel  the
solid   line  shows   our   best  relation   obtained  modifying   the
\citet{zhao09}  fitting  function  to be  valid  for  the
$c_{200}$ and the $M_{200}$ definitions, which reads as:
\begin{equation}
c_{200} = 4 \left[1 + \left( \dfrac{t_0}{3.2 t_{0.04}}\right)^8 \right]^{1/13} \,.
\label{eqct004}
\end{equation}
These  two  curves  are  also  shown in  the  other  three  panels  of
Fig.~\ref{figcmt004}.  The  different curve  types in each  panel show
equation~(\ref{eqct004})  renormalized  (see Table~\ref{tabct004})  in
order to  best-fitting  the data  points; \textbf{we  denote with  $A$ the
  re-normalization  parameter  to  fit the  corresponding  data}.   In
Table~\ref{tabct004} we summarize  the results of Fig.~\ref{figcmt004}
together with the rms with respect to the different models.

\begin{table*}
\caption{ \label{tabct004}Rms of the  measured concentration (from the
  second to the forth columns) and least-squares fit parameters of the
  data  (fifth   column)  with   respect  to   the  models   shown  in
  Fig.~\ref{figcmt004}.}
\begin{center}
\begin{tabular}{|r|c|c|c|c|c|c|}
\hline
\hline
$ $ & $|$ & $ $ & rms with respect to   & $ $ & $|$ & $ $ \\ 
model & $|$ & \citet{zhao09} & eq.~(\ref{eqct004})  & eq.~(\ref{eqct004})$\times A$ 
& $|$ & $\log(A)$ \\ \hline
$\mathrm{\Lambda}$CDM & $|$ &  0.152 & 0.143 & - & $|$ & -  \\
EXP003 & $|$ & 0.178 & 0.148 & 0.143 & $|$ & -0.04 \\ 
EXP008e3 & $|$ & 0.168 & 0.147 & 0.143 & $|$ & -0.05 \\ 
SUGRA003 & $|$ & 0.211 & 0.241 & 0.143 & $|$ & 0.19 \\  
\hline
\hline
\end{tabular}
\end{center}
\end{table*}

\begin{figure*}
\begin{center}
\includegraphics[width=7.5cm]{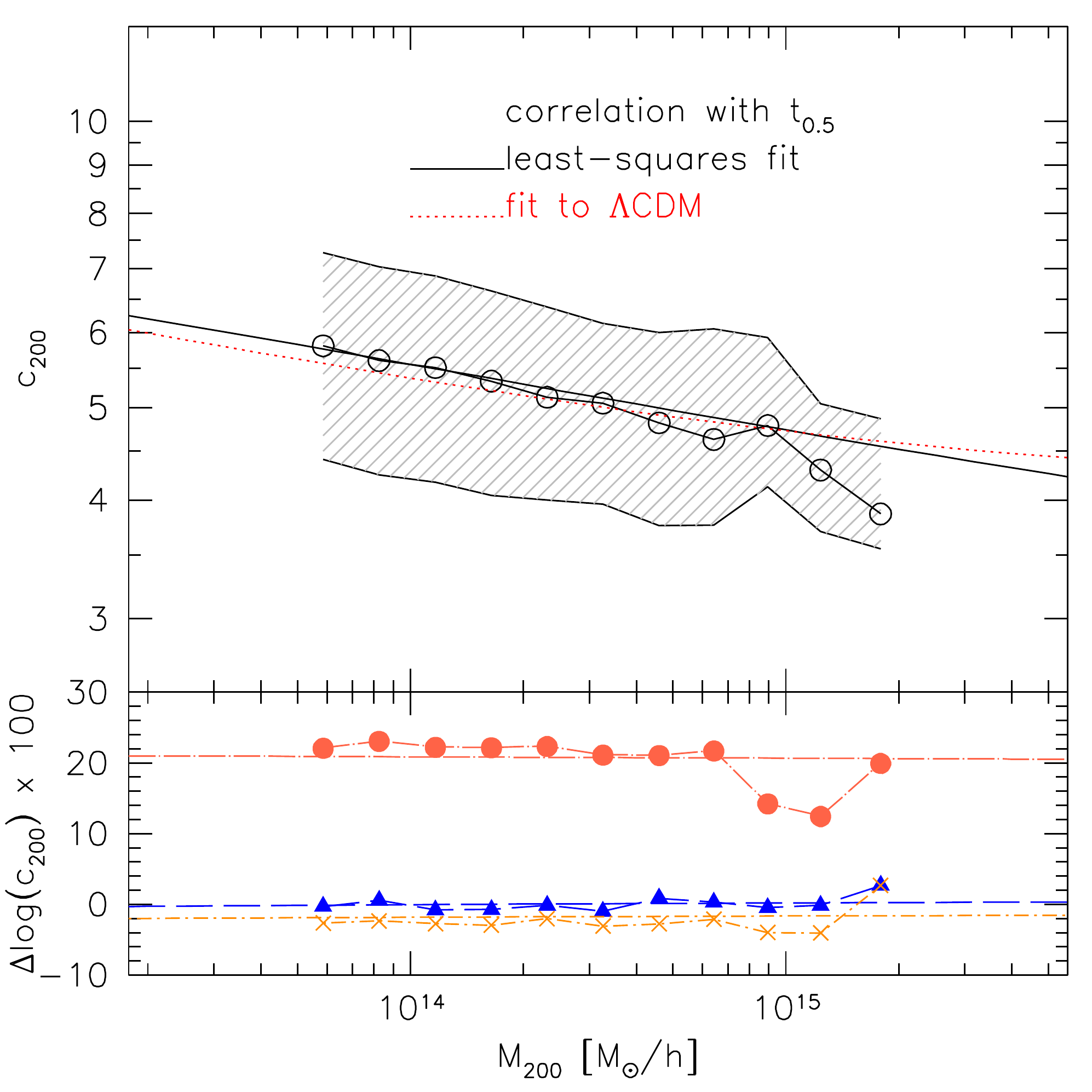}
\includegraphics[width=7.5cm]{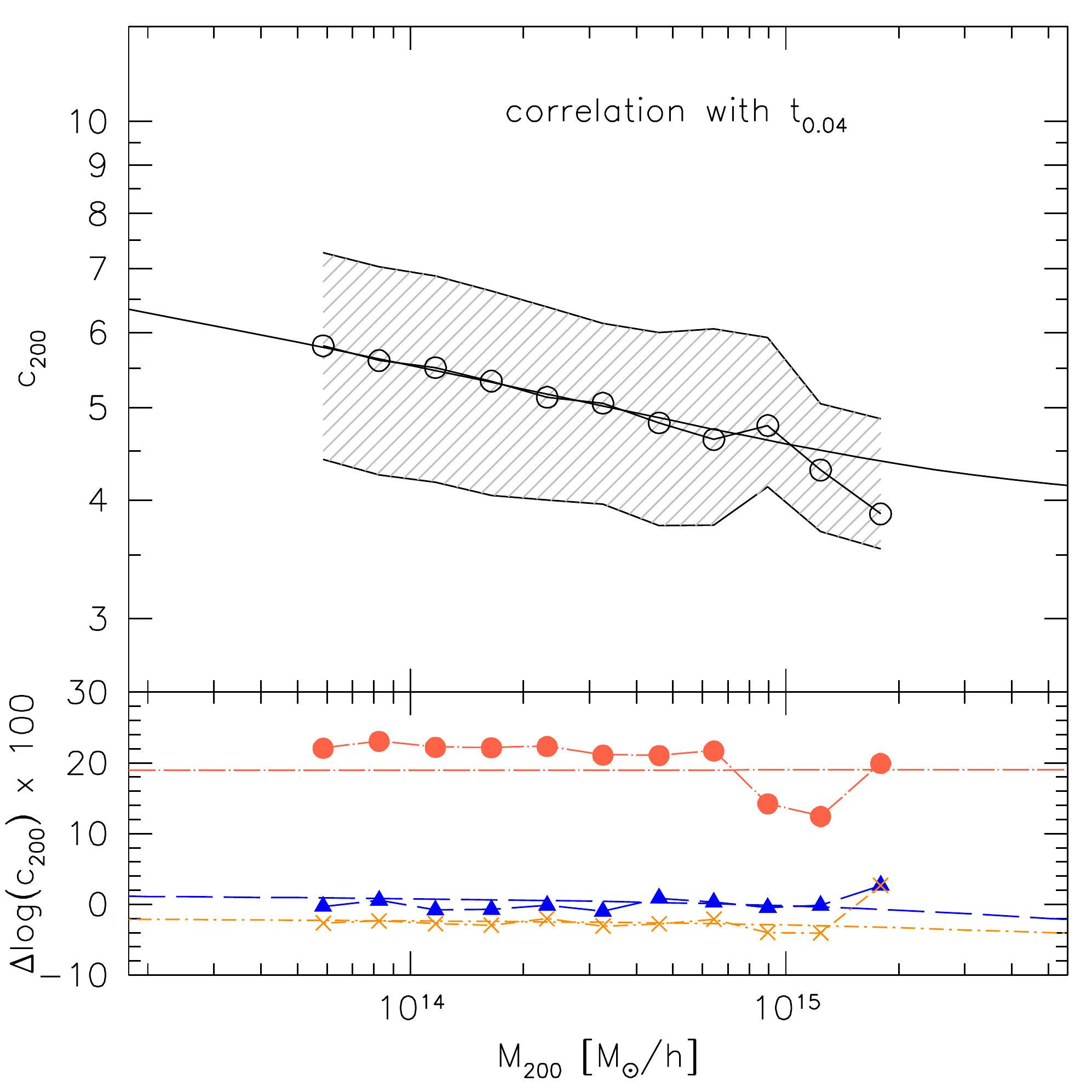}
\caption{Concentration-mass relation at $z_0=0$.  The data points show
  the  median $c-M$  relation  measured  in the  $\mathrm{\Lambda}$CDM
  simulation and  the shaded  region encloses  the quartiles.   In the
  bottom part  of each panel  the points  show the differences  of the
  $c-M$  measured in  the three  cDE simulations  with respect  to the
  $\mathrm{\Lambda}$CDM  one.  In  the  left-top panel  the two  solid
  curves show  the $c-M$ prediction using  equation~(\ref{eqct05}) and
  the least-squares relation $c_{200}-t_0/t_{0.5}$. In the bottom left
  panel the  curves show  for each  model the  residuals of  the $c-M$
  prediction  from  the  least-squares  fit  in  $c_{200}-t_0/t_{0.5}$
  relation with respect to the  one for $\mathrm{\Lambda}$CDM.  In the
  top-right     panel     the     $c-M$     prediction     is     from
  equation~(\ref{eqct004}) and in the bottom  panel the curves are the
  residuals of the predictions from renormalizing eq.~(\ref{eqct004}).
  \label{figcmrel}}
\end{center}
\end{figure*}

Using the MAH model previously  presented in this work and considering
the  correlations between  $c_{200}$ and  $t_{0.5}$ and  $c_{200}$ and
$t_{0.04}$, it is possible to estimate the concentration-mass relation
model,  in the  following way.   Given  a mass  $M_{200}$ at  redshift
$z_0$, we can estimate the redshift  at which the main halo progenitor
assembles    half    (or    $4\%$)    of    its    mass    by    using
equation~(\ref{eqmahmodz}), and then the  corresponding cosmic time by
integrating  over the  scale  factor $a=1/(1+z)$  the  inverse of  the
Hubble  constant $1/H(a)$,  from which  we can  obtain $c_{200}$.   In
Fig.~\ref{figcmrel}  we show  the median of  the concentration-mass
relation   for    all   haloes    more   massive   than    $5   \times
10^{13}M_{\odot}/h$  at redshift  $z_0=0$.   The data  points in  both
right and left panels are the  same.  In the top panels they represent
the  median  of  the   concentration-mass  relation  measured  in  the
$\mathrm{\Lambda}$CDM simulation,  and the shaded region  encloses the
first and the  third quartiles of the distribution at  fixed mass.  In
the  bottom panels  we  show  the differences  of  the  median of  the
measurements   in  the   three  cDE   models  with   respect  to   the
$\mathrm{\Lambda}$CDM one.
In  Table  \ref{tabcmrel} we  summarize  the  residuals of  the  $c-M$
estimates with respect to the different model predictions; \textbf{the
  column l-s$_{t_{0.5,\mathrm{\Lambda CDM}}}$ refers to the prediction
  made     using    the     least-squares    to     the    correlation
  $c_{200}$-$t_0/t_{0.5}$  for  the  $\mathrm{\Lambda  CDM}$  cosmology
    while  l-s$_{t_{0.5}}$  refers to  the  prediction  done with  the
    least-squares fit to the corresponding cosmology.}
\begin{table*}
\caption{ \label{tabcmrel}Rms of the measured concentration  with respect 
to the $c-M$ relation model,  as presented in the  panels of Fig.~\ref{figcmrel}.}
\begin{center}
\begin{tabular}{|r|c|c|c|c|c|c|}
\hline
\hline
$ $ & $|$ & $ $& $ $ & rms with respect to   & $ $ & \\  
model & $|$ & eq.~(\ref{eqct05})  & l-s$_{t_{0.5,\mathrm{\Lambda CDM}}}$ & l-s$_{t_{0.5}}$ & 
eq.~(\ref{eqct004}) & eq.~(\ref{eqct004})$\times A$ \\ \hline
$\mathrm{\Lambda}$CDM & $|$ &  0.153 & 0.154 & - & 0.154 & - \\
EXP003 & $|$ &  0.157 & 0.161 & 0.153 & 0.165 & 0.154 \\
EXP008e3 &$|$ &  0.156 & 0.161 & 0.152 & 0.163 & 0.151 \\
SUGRA003 &$|$ &  0.270 & 0.259 & 0.162 & 0.260 & 0.162 \\ 
\hline
\hline
\end{tabular}
\end{center}
\end{table*}

\begin{figure}
\begin{center}
\includegraphics[width=\hsize]{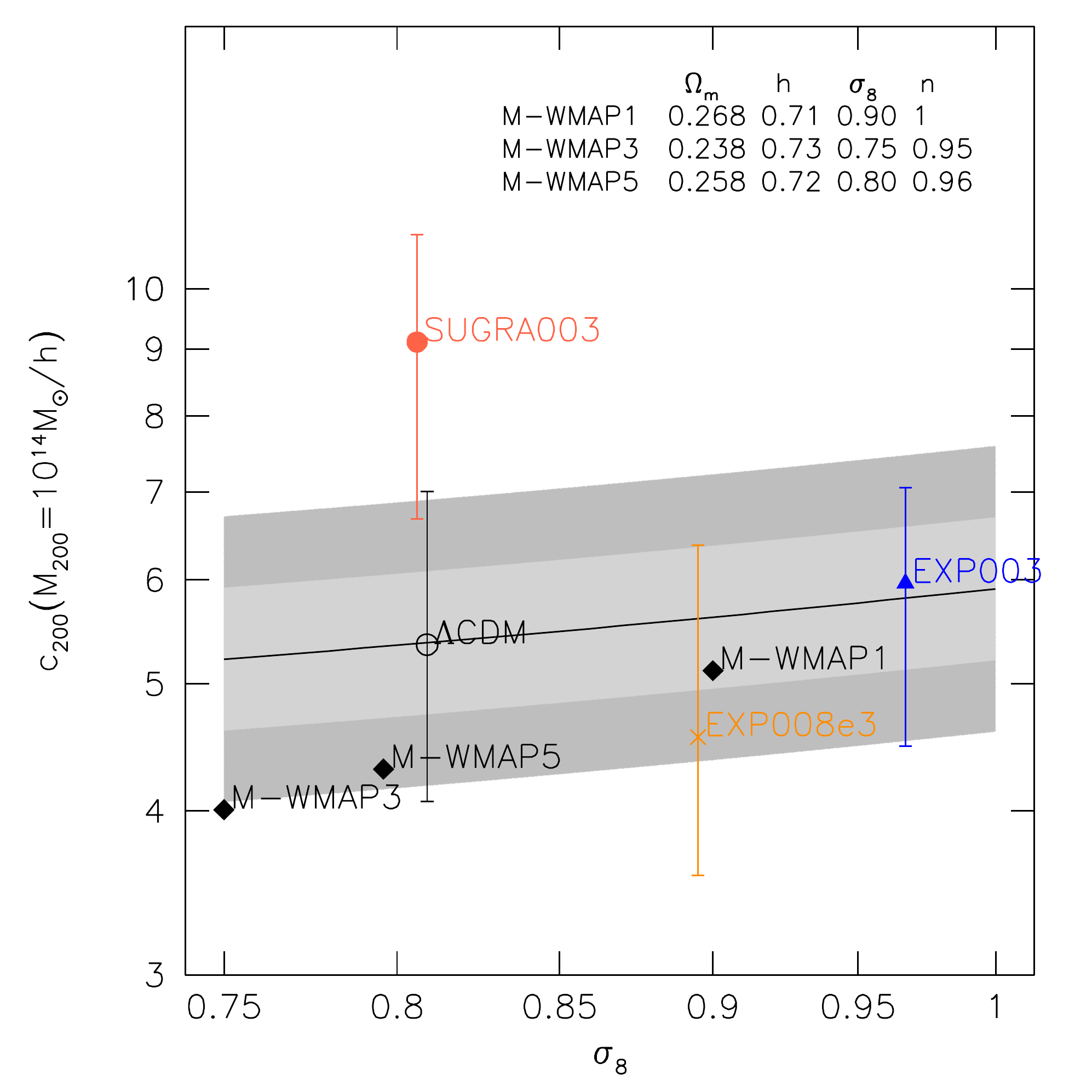}
\caption{\label{figcsigma8}Concentration-$\sigma_8$  relation   for  a
  halo  with  mass  $M_{200}=1 \times  10^{14}M_{\odot}/h$  using  the
  least-squares  fit $c_{200}-t_0/t_{0.5}$.   The colored  data points
  show  the median  of the  measurements in  the simulations  with the
  quartiles.   For  each  simulation,  we  consider  haloes  within  a
  logarithmic   mass   bin   $\mathrm{d}\log(M)=0.001$   centered   in
  $M_{200}=10^{14}M_{\odot}/h$.  The three  black data points labelled
  as M-WMAP1, M-WMAP3 and M-WMAP5 are the estimates obtained using the
  best-fitting  relation  at  $z=0$ by  \citet{maccio08}  for  the  three
  corresponding numerical simulations.}
\end{center}
\end{figure}

\begin{figure}
\begin{center}
\includegraphics[width=\hsize]{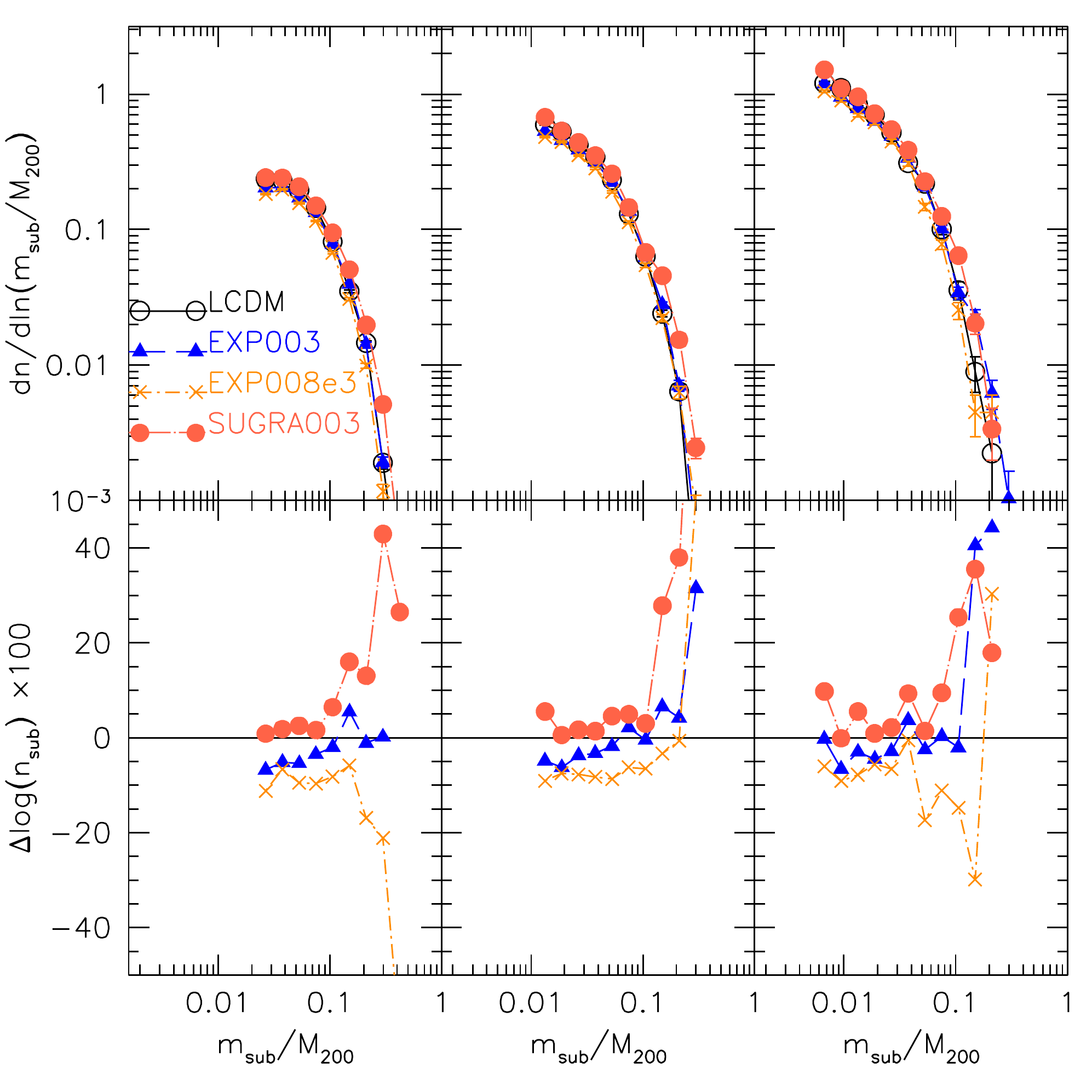}
\caption{Subhalo  mass  function  measured in  the  four  cosmological
  simulations used  in this work.   The subhalo mass is  rescaled with
  respect to the  host halo mass $M_{200}$, and for  each halo we have
  considered only subhaloes with a  distance from the host halo centre
  smaller than $R_{200}$. In the bottom part of the figure we show the
  differences  of  the  subhalo  mass function  with  respect  to  the
  measurements       done      in       the      $\mathrm{\Lambda}$CDM
  simulation.\label{figshmf}}
\end{center}
\end{figure}

\begin{figure}
\begin{center}
\includegraphics[width=\hsize]{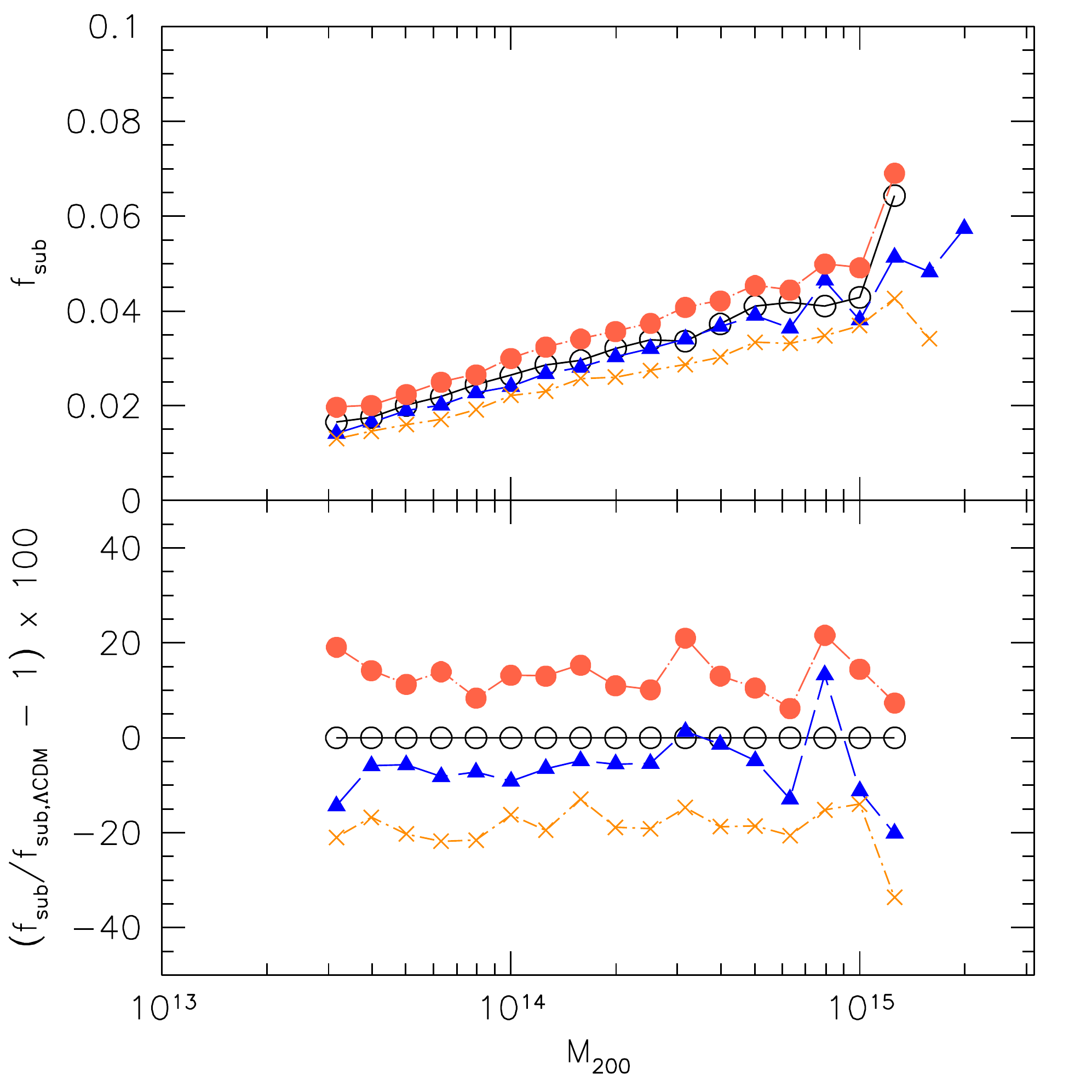}
\caption{\label{figfracmf}  Mass  fraction   in  substructures  within
  $R_{200}$ as a  function of the host halo mass.   In the bottom part
  we show the  differences of the mass fraction  in substructures with
  respect to the $\mathrm{\Lambda}$CDM measurements.}
\end{center}
\end{figure}

\citet{maccio08} and \citet{zhao09} have shown that haloes in standard
$\Lambda  $CDM  simulations  with  low  values  of  $\sigma_8$  and/or
$\Omega_m$ tend  to possess a  lower concentration due to  their later
formation epoch.  In what follows we will try to understand if this is
also the case in our non-standard cDE simulations, i.e we will enquire
whether    the    concentration    can   be    inferred    from    the
$\mathrm{\Lambda}$CDM one by simply  taking into account the different
normalization of the linear matter  power spectrum at $z=0$ within the
standard MAH model.

From the analysis  presented above, we have already  noticed that even
if  the EXP003  and EXP008e3  models have  a higher  value of  $\sigma
_{8}$, the halo concentration is  not too different from that measured
for haloes of the same  mass in the $\mathrm{\Lambda}$CDM model.  Only
the  SUGRA003  haloes  are  found   to  have  a  significantly  higher
concentration, irrespectively of their value  of $\sigma _{8}$ that is
very  similar to  the  one in  the  $\mathrm{\Lambda}$CDM run.   These
results are consistent with the previous findings of \citet{cui12}.

In order to  quantify to what extent halo  concentrations are expected
to change for different values of $\sigma _{8}$ within a $\Lambda $CDM
cosmology,     we    plot     in    Fig.~\ref{figcsigma8}    the
concentration-$\sigma_8$  prediction combining  the least-squares  fit
$c_{200}-t_0/t_{0.5}$  and  the  MAH  model   for  a  halo  with  mass
$M_{200}=1    \times     10^{14}M_{\odot}/h$    as     expected    for
$\mathrm{\Lambda}$CDM -- with the  same cosmological parameters of our
$\mathrm{\Lambda}$CDM   run   --   but  with   different   values   of
$\sigma_8$.  We  take  into  account   the  change  of  $\sigma_8$  by
renormalizing the mass variance such that:
\begin{equation}
S_{\sigma_8}(M) = \dfrac{\sigma_8^2}{0.809^2} S_{\mathrm{\Lambda CDM}}(M)\,.
\end{equation}
Fixing  all the  other  cosmological parameters,  we  notice that  the
larger $\sigma_8$ the  larger is the halo  concentration, as expected,
reflecting the fact that haloes tend to form earlier.  From the figure
we  can see  that when  $\sigma_8$  changes from  0.75 to  1 the  halo
concentration changes by $\sim 20\%$.   The data points with the error
bars show the median and the  quartiles of $c_{200}$ for the same mass
haloes  measured in  the  four  simulations at  $z=0$;  we consider  a
logarithmic   mass    bin   $\mathrm{d}\log(M)=0.001$    centered   in
$M_{200}=10^{14}M_{\odot}/h$.  The three  data points M-WMAP1, M-WMAP3
and  M-WMAP5  are  the  predictions  based  on  the  $c_{200}-M_{200}$
relations  by \citet{maccio08}  for the  three cosmological  models at
$z=0$.   The  fact   that  the  predictions  from   the  relations  by
\citet{maccio08} shift down  with respect to the  solid curve reflects
the  smaller   values  for   $\Omega_m$  adopted   their  simulations:
$\Omega_{m,\mathrm{\Lambda     CDM}}    \gtrsim\Omega_{m,M-WMA1}     >
\Omega_{m,M-WMA5}  > \Omega_{m,M-WMA3}$.   From the  figure we  notice
that  haloes  in  EXP003  possess   a  higher  concentration  than  in
$\mathrm{\Lambda}$CDM, consistently  with the higher value  of $\sigma
_{8}(z_{0})$      that      characterizes     the      model      (see
Table~\ref{tab:models}).

However,  the expected  trend of  $c_{200}$ as  a function  of $\sigma
_{8}$  is not  followed  both  in EXP008e3  and  SUGRA003. This  point
represents a central  result of the present work, and  deserves a more
detailed discussion.   It is well  known that standard cDE  models are
characterized by a suppression of the non-linear matter power spectrum
with respect  to what $\Lambda  $CDM predicts  for any given  value of
$\sigma _{8}$.  This has been  demonstrated both by running cDE N-body
simulations  with   the  same   $\sigma  _{8}$  \citep[as   done  e.g.
  in][]{Baldi_2011b}  and by  comparing  the  non-linear matter  power
spectrum extracted  from cDE N-body simulations  normalized at $z_{\rm
  CMB}\approx   1100$  to   the   predictions   of  {\small   HALOFIT}
\citep[][]{Smith_etal_2003}  for the  corresponding values  of $\sigma
_{8}$ \citep[see  e.g.][]{CoDECS}.  Such suppression is  determined by
the  action   of  the   ``friction  term"   (see  the   discussion  in
Section~\ref{secnumsim} above)  that for  standard cDE models  has the
effect   of   suppressing   the  growth   of   nonlinear   structures.
Correspondingly, for  a fixed  value of $\sigma  _{8}$ CDM  haloes are
found to  be less concentrated in  a cDE cosmology when  compared with
respect          to          $\Lambda         $CDM          \citep[see
  e.g.][]{baldi10,Baldi_2011b,Li_Barrow_2011}.     For    more
complex cDE  models, like time-dependent couplings  or the ``bouncing"
cDE  scenario, the  role  played by  the ``friction  term"  is not  so
straightforward  as  the latter  can  change  sign during  the  cosmic
evolution or have  a non-trivial interplay with the  time evolution of
the   coupling  itself.    In  such   more  general   scenarios,  halo
concentrations at  fixed $\sigma  _{8}$ can be  both higher  and lower
than in $\Lambda $CDM, depending on the specific model.

Among the models  considered in the present work,  the slight increase
of halo concentrations for the standard EXP003 cDE model shows how the
suppression associated to  the ``friction term" is  compensated by the
higher value of  $\sigma _{8}$ of this cosmological  model as compared
to $\Lambda  $CDM. Furthermore,  the lower and  higher values  of halo
concentrations  with respect  to  what is  predicted  by the  standard
$\Lambda  $CDM  scenario,  that  are observed  for  the  EXP008e3  and
SUGRA003 models, respectively, show how the dynamical evolution of the
DE scalar field can alter the structural properties of CDM haloes in a
way that  is clearly  independent of the  evolution of  linear density
perturbations.  For instance, in the case of the ``bouncing" cDE model
(SUGRA003),  halo concentrations  significantly  grow in  time at  low
redshifts \citep[as  already found by][]{cui12} due  to the particular
dynamics of the  DE scalar field that inverts its  direction of motion
at  $z_{\rm inv}\simeq  6.8$ thereby  also  changing the  sign of  the
``friction"  term  $\beta  _{c}\dot{\phi  }$  \citep[see][for  a  more
  detailed    discussion    of    the    dynamics    of    ``bouncing"
  cDE]{Baldi_2011c}.  The latter  then acts as a  dissipation term for
virialized  objects inducing  an adiabatic  contraction of  halos that
consequently evolve towards more concentrated virial configurations.

This direct relation  between the dynamics of the  underlying DE field
and the  formation and  evolution of  nonlinear structures  offers the
interesting prospect of using the  formation history of CDM haloes and
their structural  properties to disentangle cDE  effects from possible
variations of cosmological parameters  (as e.g.  $\sigma_{8}$), of the
linear galaxy  bias, and  of the mass  of cosmic  neutrinos \citep[see
  e.g.][]{LaVacca_etal_2009, marulli2011}.

\subsection{Halo substructures}

Studying the subhalo population in DM haloes extracted from a standard
cosmological     $\mathrm{\Lambda}$CDM     simulation    at     $z=0$,
\citet{giocoli10}  found that  more  concentrated  haloes, forming  at
higher redshift, tend  to possess on average  less substructures above
the same  mass ratio  $m_{sub}/M_{200}$. They  have also  stressed the
fact that  considering haloes with the  same mass, the ones  that form
earlier  not  only  possess  a  higher  concentration,  but  also  few
subhaloes. This is because halo progenitors are accreted earlier, then
spending more time in the potential well of the host halo, and thereby
tend   to   lose   a   larger   fraction   of   their   initial   mass
\citep{vandenbosch05,giocoli08b}.   To see  if  this phenomenology  is
also reflected in the cDE simulations considered in this work, we plot
in Fig.~\ref{figshmf}  the subhalo mass function  considering for each
halo all  substructures resolved  by \textsc{subfind} with  a distance
from the host  halo centre smaller than $R_{200}$. In  the top part of
the figure we show the subhalo  mass function for three different host
halo mass bin (as in Fig.~\ref{figMAH})  at $z=0$, while in the bottom
one  the  differences   with  respect  to  the   measurements  in  the
$\mathrm{\Lambda}$CDM simulation.  Since we  have shown that haloes in
SUGRA assemble  earlier and  are more concentrated  than those  in the
$\mathrm{\Lambda}$CDM run, we would have expected them to possess less
substructures, on the  contrary to what is shown in  the figure.  Also
the subhalo population in the EXP003 and EXP008e3 cosmologies does not
reflect their mass accretion history.  In these models, haloes possess
less substructures than those in $\mathrm{\Lambda}$CDM since they form
earlier,  but even  if in  EXP003 haloes  form earlier  than those  in
EXP008e3 they are  still found to host more substructures  than in the
latter   model.    This   statement   is   further   demonstrated   in
Fig.~\ref{figfracmf} where we show the halo mass fraction in subhaloes
as a function of $M_{200}$.  In the  bottom part of the figure we show
again the differences of the measurements  in the cDE with respecto to
the one in $\mathrm{\Lambda}$CDM: haloes  in SUGRA (EXP008e3) are more
(less) substructured than those in $\mathrm{\Lambda}$CDM by $\sim15\%$
while the difference between  EXP003 and $\mathrm{\Lambda}$CDM is only
of the  order of  a few  percent. We  argue that  the main  reason why
SUGRA003 (EXP008e3)  possesses more  (less) substructures at  $z=0$ is
that  halo  progenitors  at  any  redshift  $z<1$,  which  end  up  in
present-day  substructures,  are  more  (less)  concentrated  than  in
$\mathrm{\Lambda}$CDM;   so  the   gravitational  heating   and  tidal
stripping are less (more) efficient in disrupting the satellites.

\section{Summary and Conclusions}
\label{secConclusions}

Using state-of-the-art  cosmological simulations  for cDE  models from
the {\small CoDECS} Project we  have studied how collapsed haloes grow
as a function of the cosmic  time, and how present-day systems acquire
their structural properties.  We summarize  our study and main results
as follows.

\begin{itemize}

\item  We updated  the model  developed by  \citet{giocoli12b} in  the
  context  of the  extended-\citet{press74}  formalism to  be able  to
  describe the halo mass  accretion history of a $\mathrm{\Lambda}$CDM
  cosmology when haloes are identified by their $M_{200}$ mass (instead
  of $M_{vir}$), defined as the mass of the spherical region around the
  halo center enclosing  an average density 200 times  larger than the
  critical density of the Universe.

\item By rescaling the MAH model  with the normalization of the linear
  matter power  spectrum for the  different cDE runs, we  noticed that
  the simulation  results are reproduced  quite well for  small masses
  and low  redshift.  For  galaxy cluster-size haloes  and up  to high
  redshifts,  when  less  than  $1\%$   of  the  present-day  mass  is
  assembled,  the models  and  the simulation  measurements differ  by
  about  $20\%$.  Haloes  at $z=0$  in the  different cDE  cosmologies
  (standard cDE,  time-dependent cDE,  and ``bouncing"  cDE) typically
  assemble their mass  at higher redshift as compared to  those in the
  standard $\mathrm{\Lambda}$CDM run.

\item We studied the formation  redshift $z_f$ for different fractions
  $f$ of  the assembled mass:  standard and time-dependent cDE  show a
  systematically  higher   $z_f$  by   $\sim  20$  and   $\sim  10\%$,
  respectively, as  compared to  $\mathrm{\Lambda}$CDM.  On  the other
  hand  haloes  in our  ``bouncing"  cDE  cosmology have  a  formation
  redshift quite similar to  those in $\mathrm{\Lambda}$CDM, for small
  $f$, while  for large assembled  fractions $z_f$ can also  be higher
  than in $\mathrm{\Lambda}$CDM by up to $\sim 20\%$.

\item Analyzing the correlation between the halo concentration and two
  typical formation  times $t_{0.5}$  and $t_{0.04}$, we  confirm that
  the  usual   correlation  between  halo  formation   time  and  halo
  concentration at $z=0$  (the earlier the formation  time, the higher
  the concentration) still holds in cDE cosmologies.  Generalizing the
  prediction  for   $\mathrm{\Lambda}$CDM  we  provide   some  fitting
  functions for such  correlations also in cDE  models. In particular,
  in our  ``bouncing'' cDE  scenario haloes  are very  concentrated at
  $z=0$,  inconsistently with  the correlation  expected for  standard
  cosmologies \citep{cui12}.  Such structural properties could produce
  very compact and  luminous galaxies located in the  centre of haloes
  and also galaxy clusters that are very efficient for strong lensing.

\item Considering  the correlation between  $c_{200}$ and the  time at
  which the main halo progenitor assembles  for the first time half of
  its mass, we have  confirmed that for standard $\mathrm{\Lambda}$CDM
  cosmologies   the  concentration   is   a   monotonic  function   of
  $\sigma_{8}$: in  particular for a  cluster-size halo the  change in
  concentration  between  two $\mathrm{\Lambda}$CDM  cosmologies  with
  $\sigma_8 = 0.75$ and $\sigma_8$ =  1 is of about $20\%$.  Moving to
  cDE  cosmologies, we  have shown  that  only the  standard cDE  run,
  characterized  by an  exponential self-interaction  potential and  a
  constant  coupling,   is  found  to   be  in  agreement   with  such
  predictions, while  both cDE  models with a  time-dependent coupling
  and the ``bouncing''  cDE model are not found to  be consistent with
  the expected evolution of halo concentrations with $\sigma_8$.  This
  inconsistency offers  the interesting prospect of  disentangling the
  effects  of  dark  interactions   from  the  variation  of  standard
  cosmological parameters  using the evolution of  halo concentrations
  \citep{giocoli12c}.

\item The  standard mass accretion history  model is also found  to be
  not directly  applicable to  predict the  subhalo population  of the
  host  haloes at  $z =  0$  within cDE  cosmologies. In  fact, if  we
  generalize to  the cDE simulations the  statements by \citet{gao04},
  \citet{vandenbosch05}, and  \citet{giocoli08b} - valid  for standard
  $\mathrm{\Lambda}$CDM - that haloes with a higher formation redshift
  typically possess less  substructures, we would expect  to find more
  substructures in $\mathrm{\Lambda}$CDM than in all our cDE models at
  any fixed  halo mass.  On  the contrary, we  find that the  trend is
  actually reversed,  with haloes of  the same mass  having $\sim15\%$
  more substructures within $R_{200}$ in  the ``bouncing" cDE run than
  in   $\mathrm{\Lambda}$CDM,   while    haloes   forming   within   a
  time-dependent  cDE cosmology  have less  substructures by  the same
  amount, while  haloes in  standard cDE  are roughly  consistent with
  their higher formation redshift  and concentration but have slightly
  few substructures then in  $\mathrm{\Lambda}$CDM. We argue that the
  higher concentration and the larger number of clumps in the SUGRA003
  model are due to the fact  haloes formed when the average density of
  the universe was  larger, and the difference in  the average density
  between the  formation time of the  small halos and the  larger ones
  was also  larger than in  the in $\mathrm{\Lambda}$CDM.   This makes
  satellites   not   only  more   concentrated   than   in  thoes   in
  $\mathrm{\Lambda}$CDM, but  also more  concentrated relative  to the
  host halos:  the subhalos  then tend to  survive tidal  striping for
  longer time.   In summary,  while in  standard $\mathrm{\Lambda}$CDM
  simulation  -- independently  of the  small scale  behaviour of  the
  linear  power  spectrum  used  to generate  the  initial  conditions
  \citep{schneider12} --  the concentration-mass relation  is expected
  to  prove  the halo  formation  time,  the  subhalo abundance  as  a
  function of the  host halo mass validate the  dynamical friction and
  the tidal stripping \citep{vandenbosch05}.
                     
\end{itemize}
\ \\

To conclude, in the present work we have performed a detailed analysis
of  the mass  accretion history  of collapsed  haloes in  a sample  of
coupled  Dark Energy  cosmologies including  different choices  of the
Dark Energy self-interaction and  coupling functions.  We have studied
how haloes acquire  their structural properties along  the cosmic time
and  tested  whether   it  is  possible  to   attribute  the  detected
differences with respect to the standard $\mathrm{\Lambda}$CDM case to
the different  linear power spectrum normalization.  Interestingly, we
found that  this is not possible  for all coupled Dark  Energy models,
and we identified which observables  allow to break such a degeneracy.
Finally,  we have  investigated  by how  much  halo concentration  and
subhalo  abundance deviate  from $\mathrm{\Lambda}$CDM  for the  three
coupled  Dark  Energy models  included  in  our simulations  set.   In
particular,  we  showed that  the  unexpected  high concentration  and
clumpiness of  haloes make the  ``bouncing" coupled Dark  Energy model
particularly  interesting  in the  context  of  both weak  and  strong
gravitational  lensing.   We  intend   to  study  the  possibility  of
distinguishing  such models  from $\mathrm{\Lambda}$CDM  using lensing
data in future work.

\appendix 
\section{Mass Accretion History in terms of the Virial Mass for the 
  $\mathrm{\Lambda}$CDM simulation}
\label{app1}
\begin{figure*}
\begin{center}
\includegraphics[width=7cm]{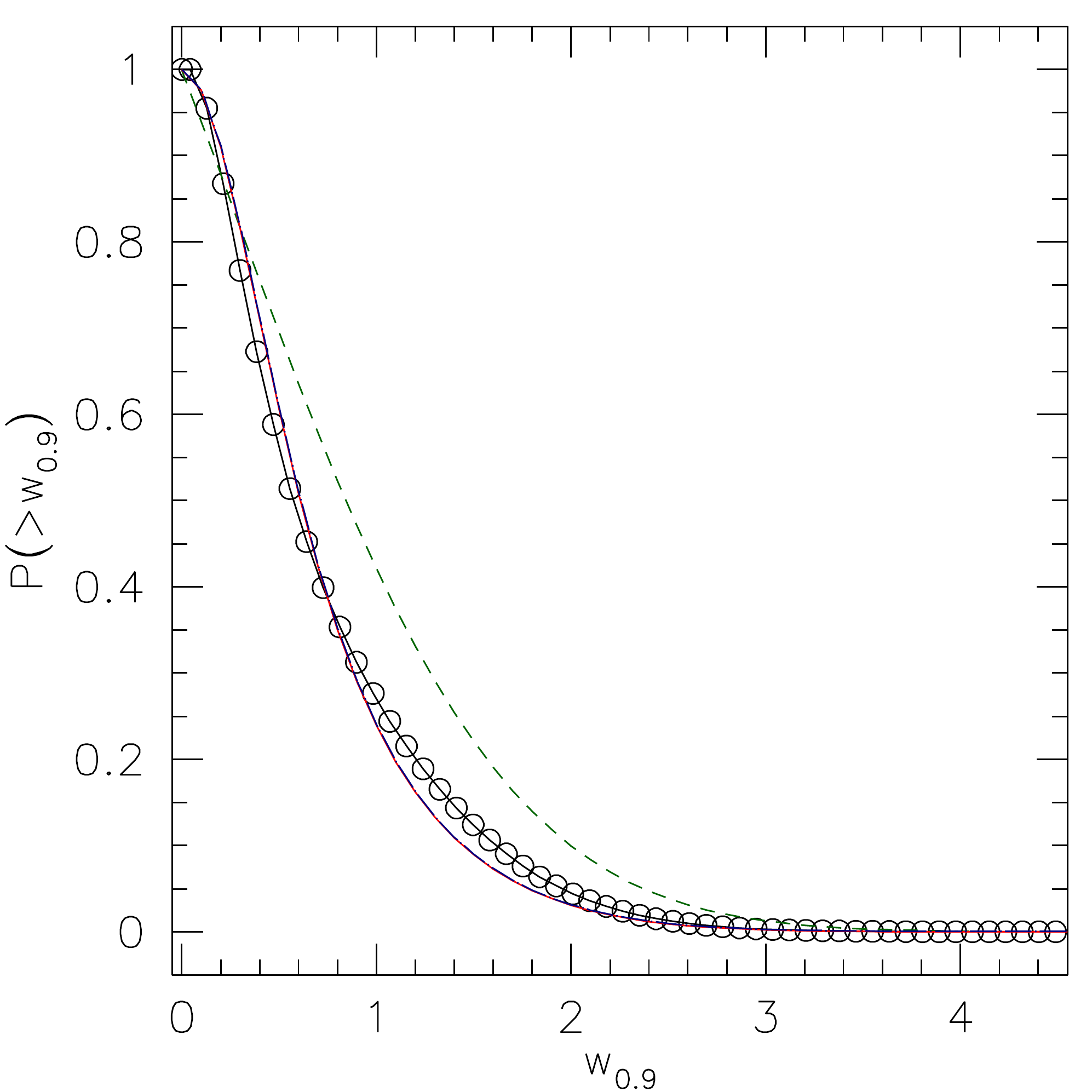}
\includegraphics[width=7cm]{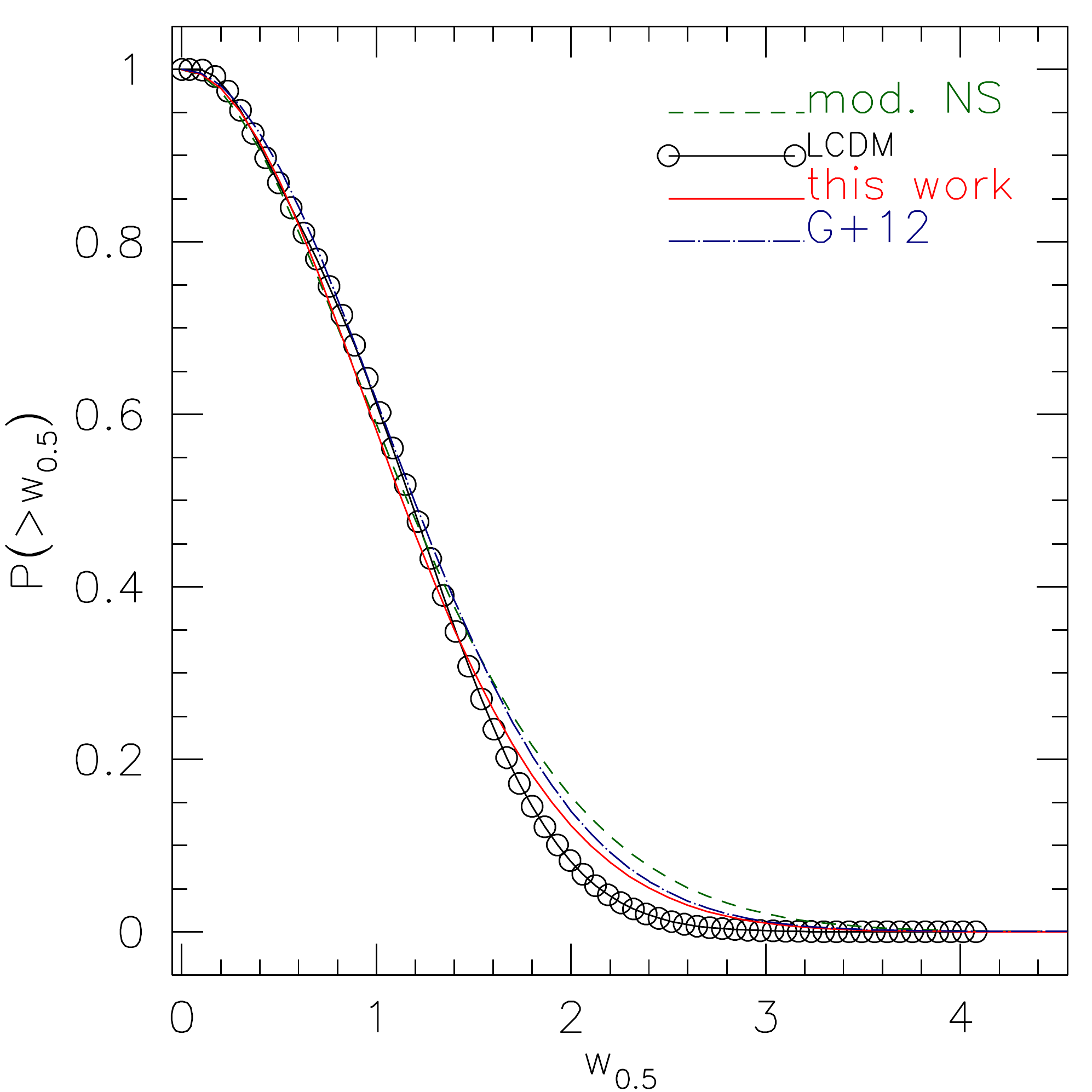}
\includegraphics[width=7cm]{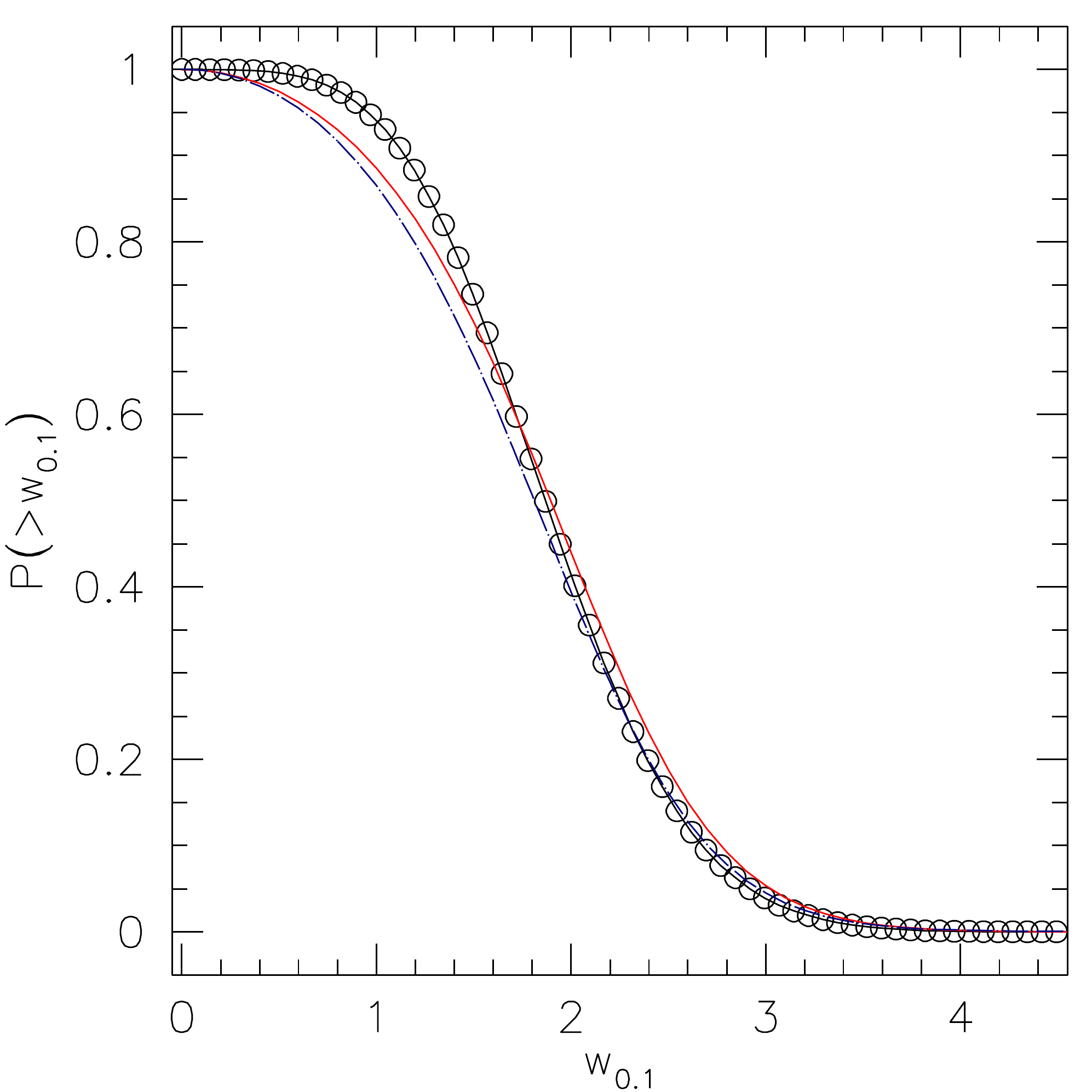}
\includegraphics[width=7cm]{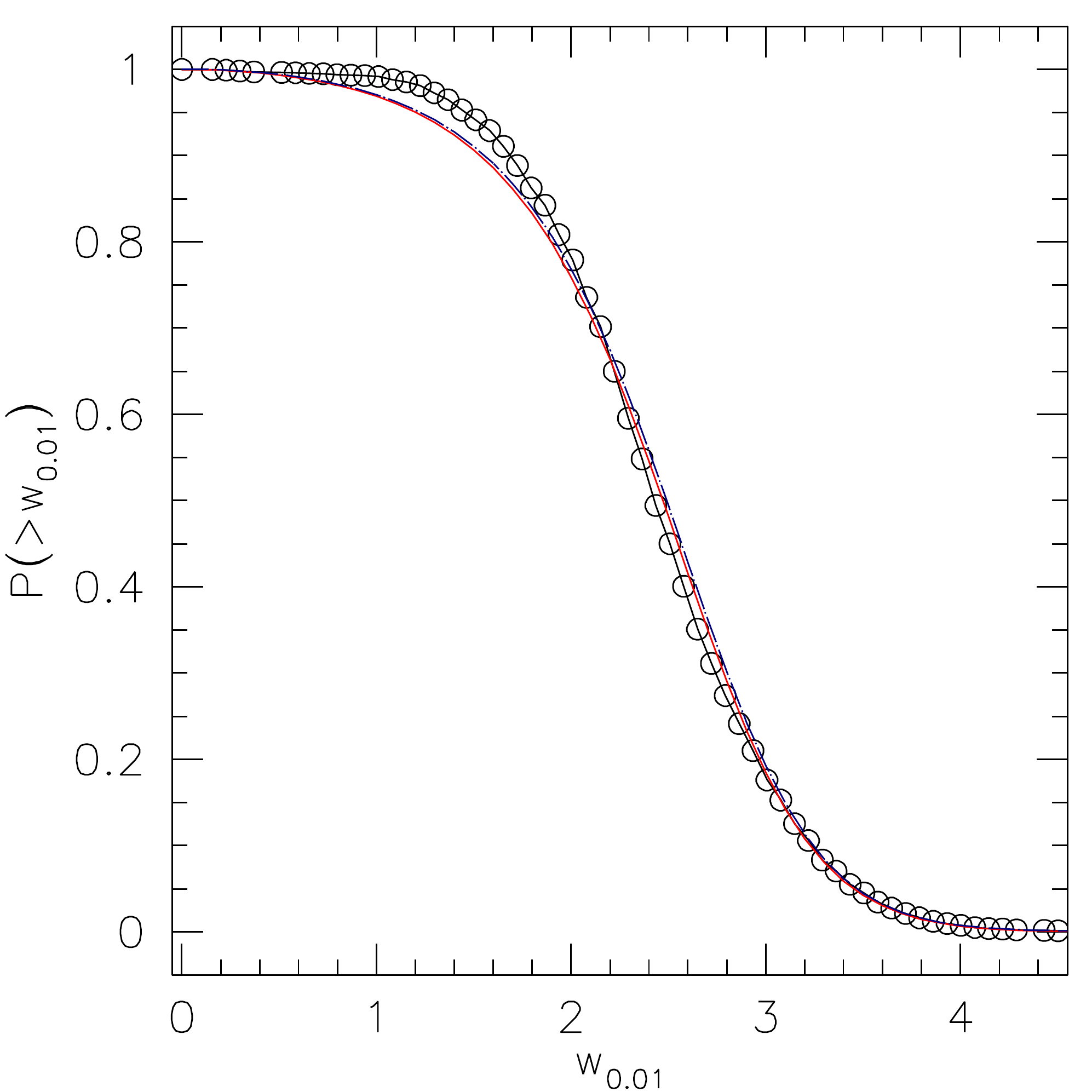}
\caption{As      Fig.~\ref{figcumPw}     but      in     terms      of
  $M_{vir}$. \label{figgiocoli12b}}
\end{center}
\end{figure*}

In this  Appendix we show that  the MAH model built  following back in
time the  haloes in the  $\mathrm{\Lambda}$CDM simulation in  terms of
their virial mass  $M_{vir}$ is in perfect agreement  with the results
obtained by \citet{giocoli12b}.  In Fig.~\ref{figgiocoli12b} we show
the  cumulative  generalized   formation  redshift  distribution  when
$90\%$, $50\%$, $10\%$ and $1\%$ of the halo mass is assembled in term
of the universal variable $w_f$. We  compute the redshift at which the
main halo  progenitor assembles a  fraction $f$  of its mass  at $z=0$
interpolating  its   mass  accretion   history  along   the  different
simulation snapshots,  and then  compute $w_f =  \left(\delta_c(z_f) -
\delta_c(z_0)\right)/\sqrt{S(f M_{vir}) - S(M_{vir})}$. The solid line
in the figure shows equation~(\ref{eqmodel1})  with the best-fitting value
of  $\alpha_f$.    The  dashed  curve   refers  to  the   relation  by
\citet{nusser99} modified  as proposed  by \citet{giocoli07a}  and the
dot-dashed  curve  equation~(\ref{eqmodel1})   with  the  $\alpha_f-f$
relation by  \citet{giocoli12b}.  We  notice that our  best-fitting  is in
very good  agreement with what  found by \citet{giocoli12b},  and does
not depend on the different  cosmological parameters of the simulation
since        $w_f$        is        a        universal        variable
\citep{press74,bond91,lacey93,lacey94}.  The small  differences can be
traced  back to  the  different code  used to  run  the two  numerical
simulations  and to  do  the post  processing  analyses.  While  HYDRA
\citep{couchman95} and  GADGET \citep{springel01a} have been  used for
the GIF2 simulation studied  by \citet{giocoli12b}, a modified version
of GADGET2 \citep{springel05a} developed to include all the additional
physical effects that characterize cDE models \citep{baldi10} has been
run four our $\mathrm{\Lambda}$CDM simulation. For the post-processing
analyses  \citet{giocoli12b}   have  adapted  and  run   the  pipeline
presented by  \citet{tormen04}, while  here we have  run the  codes by
\citet{springel01a} and \citet{boylan-kolchin09}. We recall the reader
also on  the fact  that the  GIF2 is  a pure  DM simulation  while the
{\small CoDECS} presents a baryon fluid with no hydrodynamic treatment
included.

\begin{figure}
\includegraphics[width=\hsize]{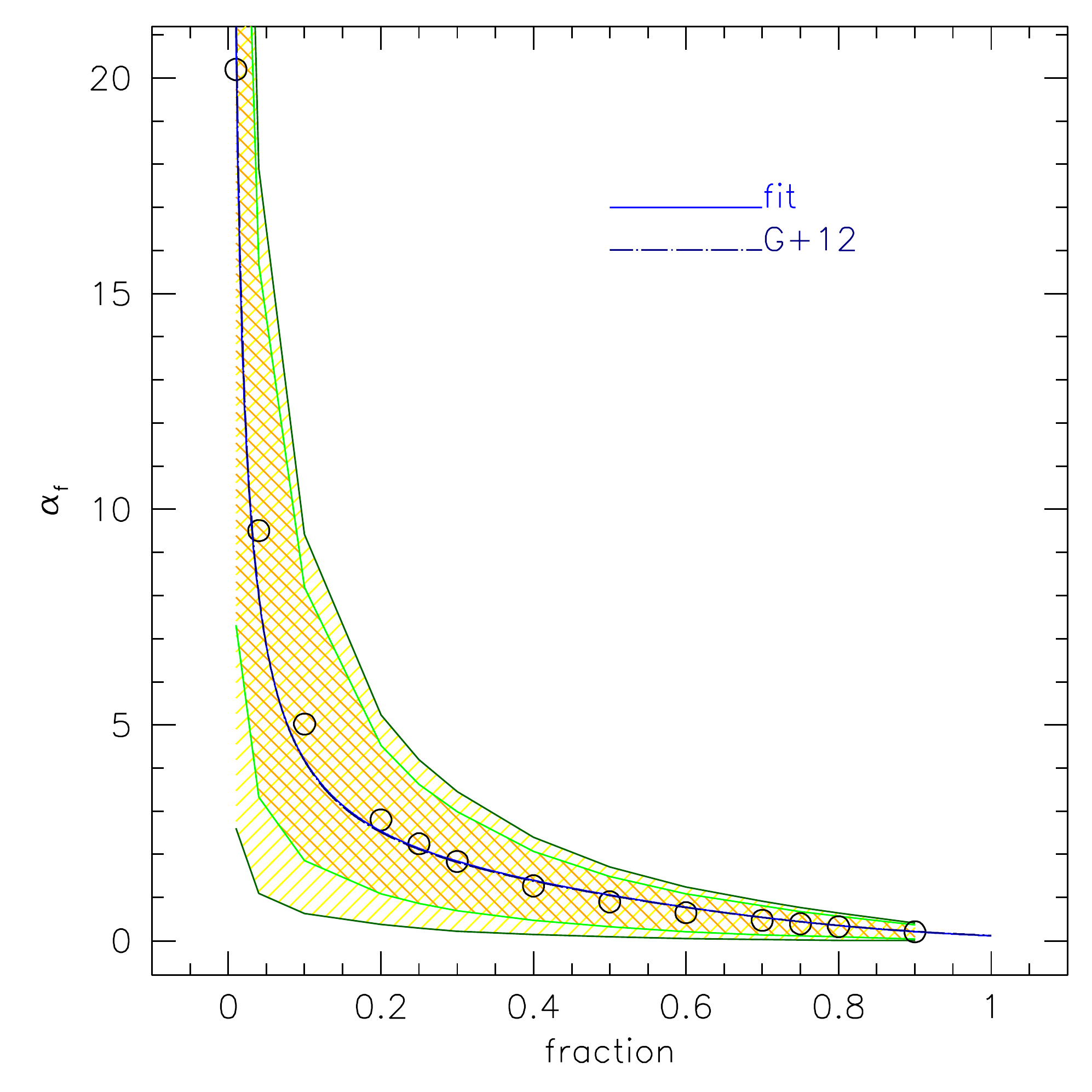}
\caption{As Fig.~\ref{figbestfit} but in terms of $M_{vir}$. \label{figMvirfit}}
\end{figure}

In Fig.~\ref{figMvirfit} we show  the correlation between  the free
parameter of the MAH model $\alpha_f$ and the assembled mass fraction.
The data  points represent  the best-fitting  values obtained  fitting the
cumulative  formation redshift  distribution for  different values  of
$f$;  the  shaded region  encloses  1$\sigma$  and 2$\sigma$  contours
around them. In  the figure, the solid curve  represents the following
relation:
\begin{equation}
 \alpha_f = \dfrac{0.837}{f^{0.7}}\, \mathrm{e}^{-2 f^3}\,,
\end{equation}
while the dot-dashed one is for:
\begin{equation}
 \alpha_f = \dfrac{0.815}{f^{0.707}}\, \mathrm{e}^{-2 f^3}\,,
\end{equation}
as   found  by   \citet{giocoli12b}.   The   two  relations   for  the
$\mathrm{\Lambda}$CDM  runs are  in perfect  agreement confirming  the
fact   that   the    MAH   model   can   be    generalized   for   any
$\mathrm{\Lambda}$CDM  cosmology,  independently of  the  cosmological
parameters.

\section{Publicly available Merger Tree files on the CoDECS database}

The merger trees  of the different {\small  L-CoDECS} simulations used
to  perform the  analysis discussed  in  the present  paper have  been
produced using a linking  algorithm outlined in \citet{Millennium} and
\citet{springel08b}.  With such algorithm, we produced a single merger
tree file for  each cosmological model of the  {\small L-CoDECS} suite
(i.e.  even  for those  models that  have not  been considered  in the
present paper).

As  an extension  of the  public {\small  CoDECS} database,  we hereby
release the merger  tree files that are now directly  available at the
{\small CoDECS} website  (http://www.marcobaldi.it/CoDECS).  These are
unformatted binary  files with  an average  size of  about 10  Gb, and
detailed instructions on how to read and  use the data can be found on
the CoDECS  guide (version 2.0)  that can also be  directly downloaded
from the {\small CoDECS} website.

The access  to these files  is subject to the  same terms of  use that
apply to the whole {\small CoDECS} public database.

\section*{acknowledgments}
We  are grateful  to  Giuseppe Tormen  and Ravi  K.   Sheth for  their
comments and suggestions.  We would  like also the thank the anonymous
referee   for  his/her   suggestions  that   helped  to   improve  the
presentation of our resuts.

CG and RBM's research is part  of the project GLENCO, funded under the
European  Seventh  Framework  Programme,  Ideas,  Grant  Agreement  n.
259349.  MB is supported by  the Marie Curie Intra European Fellowship
``SIDUN"  within the  7th European  Community Framework  Programme and
also acknowledges  support by the  DFG Cluster of  Excellence ``Origin
and  Structure  of   the  Universe''  and  by   the  TRR33  Transregio
Collaborative   Research   Network   on  the   ``DarkUniverse''.    We
acknowledge  the support  from grants  ASI-INAF I/023/12/0,  PRIN MIUR
2010-2011 ``The  dark Universe  and the  cosmic evolution  of baryons:
from current surveys to Euclid''.

\bibliographystyle{mnras}
\bibliography{cgiocoli,baldi_bibliography}
\label{lastpage}
\end{document}